\pdfoutput=1
\documentclass[sigconf]{acmart}
\usepackage[english]{babel}

\usepackage{amsmath}
\usepackage{amsfonts}
\usepackage{bbm}
\usepackage{bm}
\usepackage{mathtools}
\usepackage{scalerel} %

\usepackage{booktabs} %
\usepackage{multirow}
\usepackage{subcaption}
\usepackage{adjustbox}
\usepackage{graphicx}
\usepackage[justification=justified,skip=3pt]{caption}

\usepackage[inline]{enumitem}
\usepackage[capitalize]{cleveref}

\usepackage{tikz}
\usetikzlibrary{shapes, positioning, fit, calc}
\usetikzlibrary{arrows.meta}
\usetikzlibrary{shapes.arrows, shapes.geometric}
\usetikzlibrary{matrix}

\usepackage{pgfplots}
\usepgfplotslibrary{groupplots}
\pgfplotsset{compat=1.14}

\usepackage{sistyle}
\SIthousandsep{,}
\usepackage{soul}
\setstcolor{red}

\AtBeginDocument{%
  \providecommand\BibTeX{{%
    \normalfont B\kern-0.5em{\scshape i\kern-0.25em b}\kern-0.8em\TeX}}}

\definecolor{oyster_pink}{RGB}{238,206,205} 
\definecolor{coral_candy}{RGB}{242,208,205} 
\definecolor{baby_pink}{RGB}{246, 194, 192}
\definecolor{oyster_pink}{RGB}{238,206,205} 
\definecolor{NY_pink}{RGB}{228,136,113} 
\definecolor{petite_orchid}{RGB}{223, 157, 155}
\definecolor{carmine_pink}{RGB}{231, 76, 60}
\definecolor{deep_carmine_pink}{RGB}{236, 50, 67}
\definecolor{apricot}{RGB}{241,140,122}

\definecolor{milan}{RGB}{255, 254, 169}
\definecolor{casablanca}{RGB}{244, 178, 84}
\definecolor{texas}{RGB}{245, 232, 123}
\definecolor{maize}{RGB}{249, 212, 156}

\definecolor{double_pearl_lusta}{RGB}{253, 242, 208}

\definecolor{oasis}{RGB}{253, 242, 208}
\definecolor{linen}{RGB}{251, 239, 227}

\definecolor{zanah}{RGB}{220, 233, 213}
\definecolor{frostee}{RGB}{217, 231, 214}
\definecolor{norway}{RGB}{158, 194, 132}
\definecolor{malibu}{RGB}{110, 180, 240}

\definecolor{link_water}{RGB}{221, 232, 250}

\definecolor{spring_leaves}{RGB}{46, 83, 117}

\definecolor{venice_blue}{RGB}{87, 135, 105}
\definecolor{boston_blue}{RGB}{68, 147, 161}

\definecolor{napa}{RGB}{163, 154, 137}

\definecolor{mexican_red}{RGB}{170, 41, 37}
\definecolor{valencia}{RGB}{214, 87, 70}

\definecolor{riptide}{RGB}{141,211,199}
\definecolor{pale_prim}{RGB}{255,255,179}
\definecolor{lavender_gray}{RGB}{190,186,218}
\definecolor{salmon}{RGB}{242,131,107}
\definecolor{seagull}{RGB}{128,177,211}
\definecolor{rajah}{RGB}{253,180,98}
\definecolor{yellow_green}{RGB}{198,222,119}
\definecolor{classic_rose}{RGB}{252,205,229}
\definecolor{feijoa}{RGB}{178,223,138}

\definecolor{cruise}{RGB}{179,226,205}

\definecolor{periwinkle}{RGB}{203,213,232}
\definecolor{snow_flurry}{RGB}{230,245,201}
\definecolor{buttermilk}{RGB}{255,242,174}

\definecolor{sundown}{RGB}{249, 180, 181}
\definecolor{spindle}{RGB}{179,205,227}
\definecolor{tea_green}{RGB}{204,235,197}
\definecolor{languid_lavender}{RGB}{222,203,228}
\definecolor{champagne}{RGB}{254,217,166}
\definecolor{cream}{RGB}{255,255,204}

\definecolor{monte_carlo}{RGB}{135,204,194}
\definecolor{melon}{RGB}{254,191,181}
\definecolor{granny_smith_apple}{RGB}{150,214,150}
\definecolor{watusi}{RGB}{254,221,207}
\definecolor{see_green}{RGB}{161,228,195}

\definecolor{moss_green}{RGB}{170,216,176}
\definecolor{opal}{RGB}{164,207,190}

\definecolor{pale_turquoise}{RGB}{172,240,242}
\definecolor{Madang}{RGB}{190,235,159}
\definecolor{pixie_green}{RGB}{183,214,170}
\definecolor{coral_andy}{RGB}{243,204,205}
\definecolor{manhattan}{RGB}{226,180,125}
\definecolor{quartz}{RGB}{219,223,238}
\definecolor{spring_sun}{RGB}{242,243,195}
\definecolor{dairy_cream}{RGB}{254,226,189}
\definecolor{surf_crest}{RGB}{205,230,208}
\definecolor{french_pass}{RGB}{195,232,246}
\definecolor{cosmos}{RGB}{248,209,210}
\definecolor{portafino}{RGB}{245,237,160}
\definecolor{sail}{RGB}{163,205,235}
\definecolor{hint_green}{RGB}{226,246,209}

\definecolor{jet_stream}{RGB}{188, 214, 210}

\definecolor{azalea}{RGB}{251, 196, 196}
\definecolor{wewak}{RGB}{244, 143, 150}
\definecolor{bittersweet}{RGB}{255,111,105}
\definecolor{sunset_orange}{RGB}{242,89,75}
\definecolor{light_coral}{RGB}{244, 127, 123}
\definecolor{carnation}{RGB}{245, 80, 86}
\definecolor{flamingo}{RGB}{237, 88, 85}

\definecolor{fire_engine_red}{RGB}{210,44,41}
\definecolor{amaranth}{RGB}{234,46,73}
\definecolor{ku_crimson}{RGB}{243, 0, 25}
\definecolor{fire_engine_red}{RGB}{206, 37, 51}
\definecolor{copper_rust}{RGB}{155, 64, 74}

\definecolor{chilean_fire}{RGB}{215, 87, 44}

\definecolor{japanese_laurel}{RGB}{53, 116, 40}

\definecolor{turmeric}{RGB}{211, 178, 76}
\definecolor{saffron}{RGB}{249,193,62}
\definecolor{my_sin}{RGB}{255, 176, 59}
\definecolor{tree_poppy}{RGB}{246, 154, 27}
\definecolor{jaffa}{RGB}{240, 131, 58}
\definecolor{crusta}{RGB}{254, 127, 44}
\definecolor{tahiti_gold}{RGB}{223, 102, 36}
\definecolor{outrageous_orange}{RGB}{255, 100, 45}
\definecolor{safety_orange}{RGB}{254, 106, 0}

\definecolor{turquoise}{RGB}{41,217,194}
\definecolor{puerto_rico}{RGB}{94, 194, 166}
\definecolor{mountain_meadow}{RGB}{0, 163, 136}
\definecolor{free_speech_aquamarine}{RGB}{0, 156, 114}
\definecolor{java}{RGB}{2,190,196}

\definecolor{matisse}{RGB}{25, 104, 167}
\definecolor{shakespeare}{RGB}{85, 154, 193}
\definecolor{mona_lisa}{RGB}{246,152,134}

\definecolor{bgc}{RGB}{245,245,245}
\definecolor{tuatara}{RGB}{67, 67, 67}
\definecolor{aluminum}{RGB}{153,153,153}
\definecolor{silver}{RGB}{191,191,191}
\definecolor{platinum}{RGB}{228,228,228}
\definecolor{mercury}{RGB}{230,230,230}
\definecolor{gallery}{RGB}{240,240,240}
\definecolor{athens_gray}{RGB}{236, 240, 241}
\definecolor{ship_gray}{RGB}{77,77,77}

\definecolor{early_dawn}{RGB}{252,243,218}
\definecolor{egg_shell}{RGB}{238, 234, 215}
\definecolor{midnight}{RGB}{0, 29, 50}
\definecolor{sundown}{RGB}{249, 180, 181}
\definecolor{sun_shade}{RGB}{255, 144, 68}
\definecolor{sushi}{RGB}{117, 168, 47}
\definecolor{tomato}{RGB}{255, 97, 56}
\definecolor{ice_cold}{RGB}{169,232,220}

\definecolor{jelly_bean}{RGB}{45, 126, 150}
\definecolor{celestial_blue}{RGB}{52, 152, 219}
\definecolor{curious_blue}{RGB}{41, 128, 185}
\definecolor{french_blue}{RGB}{0, 112, 182}
\definecolor{matisse}{RGB}{25, 104, 167}

\definecolor{biscay}{RGB}{44, 62, 80}

\definecolor{cosmic_latte}{RGB}{222, 247, 229}
\definecolor{chinook}{RGB}{163, 232, 178}
\definecolor{padua}{RGB}{121, 189, 143}
\definecolor{ocean_green}{RGB}{79, 176, 112}
\definecolor{pastel_green}{RGB}{107, 227, 135}
\definecolor{chateau_green}{RGB}{69, 191, 85}
\definecolor{RoyalBlue}{RGB}{69, 191, 85}
\definecolor{pigment_green}{RGB}{0, 175, 79}
\definecolor{fern}{RGB}{101,197,117}
\definecolor{killarney}{RGB}{56, 113, 66}
\definecolor{viridian}{RGB}{70, 137, 102}

\AtBeginDocument{%
  \providecommand\BibTeX{{%
    \normalfont B\kern-0.5em{\scshape i\kern-0.25em b}\kern-0.8em\TeX}}}

\setcopyright{acmcopyright}
\copyrightyear{2021}
\acmYear{2021}
\acmDOI{10.1145/1122445.1122456}

\acmConference[SIGIR '21]{Proceedings of the 44th International ACM SIGIR Conference on Research and Development in Information Retrieval (SIGIR'21)}{July 11--15, 2021}{MONTRÉAL, Canada}
\acmBooktitle{SIGIR '21: Proceedings of the 44th International ACM SIGIR Conference on Research and Development in Information Retrieval (SIGIR'21),
  July 11--15, 2021, MONTRÉAL, Canada}
\acmPrice{15.00}
\acmISBN{978-1-4503-XXXX-X/18/06}

\begin{document}

\title{Contrastive Learning for Sequential Recommendation}

\author{Xu Xie$^1$,
        Fei Sun$^2$,
        Zhaoyang Liu$^2$,
        Shiwen Wu$^1$,
        Jinyang Gao$^2$,
        Bolin Ding$^2$,
        Bin Cui$^1$
        }
\affiliation{\institution{$^1$School of EECS \& Key Laboratory of High Confidence Software Technologies (MOE), Peking University\\
                          $^2$Alibaba Group
                          }
            }
\email{{xu.xie, wushw.18, bin.cui}@pku.edu.cn, {ofey.sf, jingmu.lzy, jinyang.gjy, bolin.ding}@alibaba-inc.com
}

\begin{abstract}
Sequential recommendation methods play a crucial role in modern recommender systems because of their ability to capture a user's dynamic interest from her/his historical interactions.
Despite their success, we argue that these approaches usually rely on the sequential prediction task to optimize the huge amounts of parameters.
They usually suffer from the data sparsity problem, which makes it difficult for them to learn high-quality user representations.
To tackle that, inspired by recent advances of contrastive learning techniques in the computer version, 
we propose a novel multi-task model called \textbf{C}ontrastive \textbf{L}earning for \textbf{S}equential \textbf{Rec}ommendation~(\textbf{CL4SRec}).
CL4SRec not only takes advantage of the traditional next item prediction task but also utilizes the contrastive learning framework to derive self-supervision signals from the original user behavior sequences.
Therefore, it can extract more meaningful user patterns and further encode the user representation effectively.
In addition, we propose three data augmentation approaches to construct self-supervision signals.
Extensive experiments on four public datasets demonstrate that CL4SRec achieves state-of-the-art performance over existing baselines by inferring better user representations.
\end{abstract}



\maketitle

\section{Introduction}

Recommender systems have been widely employed in online platforms, e.g., Amazon and Alibaba, to satisfy the users’ requirements.
On these platforms, users’ interests hidden in their behaviors are intrinsically dynamic and evolving over time, which makes it difficult for platforms to make appropriate recommendations.
To cope with this problem, various methods have been proposed to make \textit{sequential recommendations} by capturing users’ dynamic interests from their historical interactions~\cite{GRU4Rec,FPMC,caser,SASRec,Bert4rec}. 

For the sequential recommendation task, the essential problem is how to infer a high-quality representation for each user through their historical interactions.
With these user representations, we can easily recommend suitable items for each user.
Thus, the mainline of research work seeks to derive better user representation by using more powerful sequential models.
Recently, with the advances of deep learning techniques,  
a lot of works employ deep neural networks to handle this problem and obtain significant performance improvements~\cite{GRU4Rec,SASRec,caser,SRGNN}.
These sequential models, such as Recurrent Neural Network (RNN)~\cite{RNN} and Self-attention~\cite{Transformer}, can learn effective representations of users' behaviors by capturing more complicated sequential patterns.
Some previous works also adopt Graph Neural Networks (GNN)~\cite{SRGNN,FGNN} to explore more complex item transition patterns in user sequences.
Although these methods have achieved promising results, they usually only utilize the item prediction task to optimize these huge amounts of parameters, which is easy to suffer from data sparsity problem~\cite{Google,S3}.
When the training data is limited, these methods may fail to infer appropriate user representations.

Recently, self-supervised learning techniques have made a great breakthrough for the representation learning in domains like computer vision (CV) and natural language processing (NLP)~\cite{BERT,GPT-2,Simclr,Jigsaw}.
They attempt to extract intrinsic data correlation directly from unlabeled data.
Inspired by the successes of self-supervised learning, we aim to use self-supervised learning techniques to optimize the user representation model for improving sequential recommender systems.
To achieve this goal, one straight-forward way could be to directly adopt a powerful sequential model like GPT~\cite{GPT-2} on a larger user behavior corpus.
However, this way is not suitable for the recommender system for two reasons.
\begin{enumerate*}[label=(\roman*)]
\item Recommender systems usually do not have a larger corpus for pre-training.
Meanwhile, different from the NLP area, different tasks in the recommender systems usually do not share the same knowledge, thus restricting the application of pre-training.
\item The objective function of such predictive self-supervised learning is almost the same as the goal of sequential recommendation, which is often modeled as a sequential prediction task.
Applying another same objective function on the same data cannot help the user representation learning in the sequential recommendation.
\end{enumerate*}

Due to the issues mentioned before, the application of self-supervised learning in recommendation systems is less well studied.
The closest line of research seeks to enhance feature representations via self-supervision signals derived from the intrinsic structure of the raw feature data, e.g., item attributes~\cite{Google,S3}.
These previous works usually focus on improving the representations at the item level.
However, how to acquire accurate representations in the user behavior sequence level is not well studied.

Different from the previous study focusing on enhancing item representation through the self-supervised task on item features, we aim to learn better sequence representations via self-supervised signals on the user behavior sequence level, even with only identifier (ID) information.
Specifically, we propose a novel model called \textbf{C}ontrastive \textbf{L}earning for \textbf{S}equential \textbf{Rec}ommendation (CL4SRec).
Our model combines the traditional sequential prediction objective with the contrastive learning objectives. 
With the contrastive learning loss, we encode the user representation by maximizing the agreement between differently augmented views of the same user interaction sequence in the latent space. 
In this way, CL4SRec can infer accurate user representations and then easily select appealing items for each user individually. 
Furthermore, we propose three data augmentation methods (crop/mask/reorder) to project user interaction sequences to different views.
We conduct extensive experiments on four real-world public recommendation datasets.
Comprehensive experimental results verify that CL4SRec achieves state-of-the-art performance compared with competitive methods. %

Our primary contributions can be summarized as follows:
\begin{itemize}
    \item We propose a novel model called Contrastive Learning for Sequential Recommendation (CL4SRec), which infers accurate user representations with only users' interaction behaviors. 
    To the best of our knowledge, this is the first work to apply contrastive learning to the sequential recommendation.
    \item We propose three different data augmentation approaches, including cropping, masking, and reordering, to construct different views of user sequences. 
    \item Extensive experiments on four public datasets demonstrate the effectiveness of our CL4SRec model. 
    Compared with all competitive baselines, 
    the improvements brought by the contrastive learning framework is almost 7.37\% to 11.02\% according to the ranking metric on average. 
\end{itemize}

\section{Related Work}
\subsection{Sequential Recommendation}

Early works on sequential recommendation usually model the sequential patterns with the Markov Chain~(MC) assumption. 
\citet{FPMC} combined the first-order MC and matrix factorization and achieved promising results in capturing the subsequent action. %
To consider more preceding items and extract more complicated patterns, the high-order MC was also explored in the following work~\cite{Fossil}.
With the recent advance of deep learning, 
Recurrent Neural Networks~(RNN)~\cite{RNN}, and its variants, such as Long Short-Term Memory~(LSTM)~\cite{LSTM}, and Gated Recurrent Unit~(GRU)~\cite{GRU4Rec,UserGRU}, are applied for user behavior sequences.
For example, \citeauthor{GRU4Rec}~\cite{GRU4Rec} adopt GRU modules to the session-based recommendation task with the ranking loss function.
The following variants modified GRU4Rec by introducing attention mechanism~\cite{NARM,AIR,SDM}, hierarchical structure~\cite{Hierarchical}, and user-based gated network~\cite{UserGRU}.

Except for recurrent neural networks, other deep learning models are also adopted for sequential recommendation tasks and achieved excellent performance. 
\citet{caser} propose a Convolutional Sequence Embedding Recommendation Model~(Caser).
By abstracting the user behaviors as a one-dimensional image, the horizontal and vertical convolutional filters are effective in capturing the sequential patterns in neighbors.
\citet{SUM} and \citet{Memory} leverage the memory-augmented neural network to store and update useful information explicitly.
The self-attention mechanism~\cite{Transformer} has shown promising potential in modeling sequential data recently.
\citet{SASRec} utilize Transformer layers to assign weights to previous items adaptively.
\citet{Bert4rec} improve that by adopting a bidirectional Transformer to incorporate user behaviors information from both directions, 
since the user’s historical interactions may not be a rigid order~\cite{Rigidorder}.

Along another line, graph neural network (GNN)~\cite{GNN} has aroused comprehensive attention in the deep learning society.
Since GNN can capture complicated item transition patterns hidden in user sequences, many works attempt to utilize powerful GNN techniques in sequential recommendation tasks, especially when each user sequence can be constructed as a graph.
For example, \citet{SRGNN} adopt the gated graph network to aggregate the neighborhood information and make the adjustment for the directed graph.
\citet{FGNN} utilize the attention mechanism as the aggregation function to differentiate the influence of different neighbors instead of mean pooling.
To capture both global and local structure, ~\citet{GCSAN} combine self-attention with GNN and achieve promising results for the session-based recommendation.

\subsection{Self-supervised Learning}
Recently, Various self-supervised learning methods have shown their superior ability on learning useful representations from unlabeled data.
Their basic idea is to construct training signals from the raw data with designed pretext tasks to train the model.
We categorize the existing pretext tasks into two categories, prediction task and discrimination task.

The prediction tasks have been widely adopted in CV and NLP areas, which usually transform the raw data into another form and then predict the transformation details.
In the CV community, various self-supervised learning pretext tasks are proposed, including predicting image rotations~\cite{Image_rotation} and relative patch locations~\cite{relative_patch}, solving jigsaw puzzles~\cite{Jigsaw}, and colorization problems~\cite{Colorization}.
For NLP, most works are focusing on acquiring universal word representations and effective sentence encoder through self-supervised learning.
Language model is naturally a self-supervised objective that learns to predict the next word given the previous word sequence~\cite{Language_Model}.
Besides, various pretext tasks have also been proposed recently, e.g., the Cloze task~\cite{BERT} and restoring sentence order~\cite{lewis:acl2020:bart}. %
\tikzset{
  FARROW/.style={arrows={-{Latex[length=1.25mm, width=1.mm]}}},
  DFARROW/.style={arrows={{Latex[length=1.25mm, width=1.mm]}-{Latex[length=1.25mm, width=1.mm]}}},
  origin/.style = {circle, fill=monte_carlo, minimum width=1.8em, align=center, inner sep=0, outer sep=0, font=\scriptsize},
  aug1/.style = {circle, fill=flamingo!72, minimum width=1.8em, align=center, inner sep=0, outer sep=0, font=\scriptsize},
  aug2/.style = {circle, fill=casablanca, minimum width=1.8em, align=center, inner sep=0, outer sep=0, font=\scriptsize},
  encoder/.style = {rectangle, fill=Madang!82, minimum width=11em, minimum height=3em, align=center, rounded corners=3},
  emb_layer/.style = {rectangle, fill=languid_lavender!72, minimum width=11em, minimum height=2em, align=center, rounded corners=3},
  project/.style = {rectangle, fill=hint_green, minimum width=7em, minimum height=2.4em, align=center, rounded corners=2},
  v_rep/.style={
       rectangle split,
       rectangle split part align=base,
       rectangle split horizontal=true,
       rectangle split draw splits=true,
       rectangle split parts=7,
       rectangle split part fill={red!30, blue!20, athens_gray!80, matisse, silver},
       draw=gray, %
       very thin,
       minimum height=2em,
       minimum width=3em,
       inner sep=2.5pt,
       text centered,
       text=gray,
       rounded corners=1
       },
}

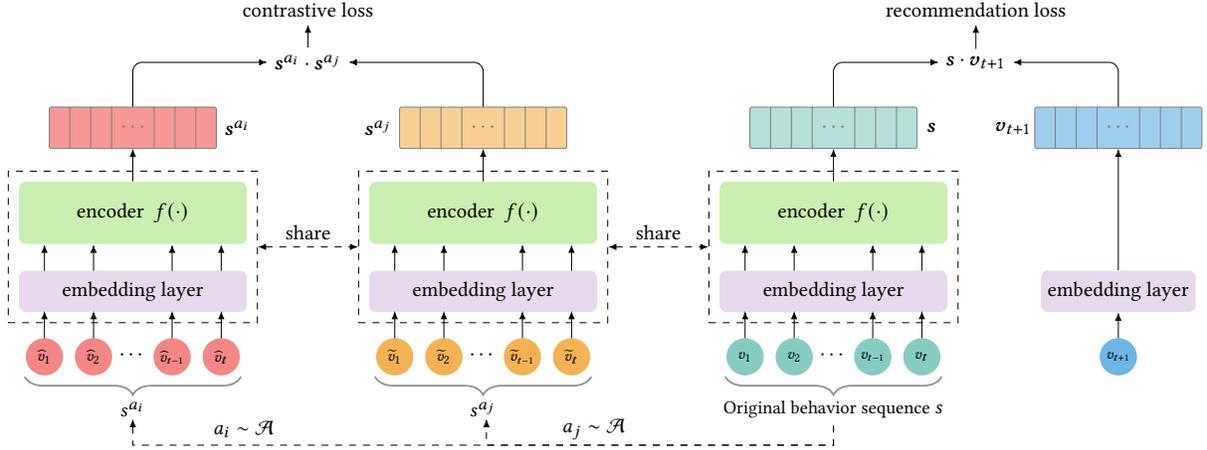
\begin{figure*}
\resizebox{0.9\linewidth}{!}{
    \begin{tikzpicture}

    \node[] (o) at (0, 0) {};
    
    \node [above of=o, node distance=0.8cm, align=center] (v_dots) {$\cdots$};
    \node [origin, left of=v_dots, node distance=0.6cm] (v2) {$v_2$};
    \node [origin, left of=v2, node distance=0.75cm] (v1) {$v_1$};
    \node [origin, right of=v_dots, node distance=0.6cm, inner sep=0, outer sep=0] (v4) {$v_{\scaleto{t-1}{2.8pt}}$};
    \node [origin, right of=v4, node distance=0.75cm] (v5) {$v_t$};
    
    \node [emb_layer, above of=v_dots, node distance=1cm] (o_emb) {embedding layer};
    \node [encoder, above of=o_emb, node distance=1.2cm] (o_enc) {encoder~ $f(\cdot)$};
     \node [dashed, draw=black, fit={([yshift=-0.5mm, xshift=-0.5mm] o_emb.south west) ([yshift=0.5mm, xshift=0.5mm] o_enc.north east)}, inner sep=3] (o_m) {};
     
      \node[v_rep, rectangle split part fill={monte_carlo!62}, above of=o_enc, node distance=1.3cm] (v_i) {
      \nodepart{four} $\,\cdots$
     };
    
    \draw [thick,draw=black!42, decorate,decoration={brace,amplitude=8pt,mirror}] ([xshift=-1mm, yshift=-1mm] v1.south west) -- ([xshift=1mm, yshift=-1mm] v5.south east) node[midway,yshift=-0.5cm] (o_a) {{\small Original behavior sequence} $s$};

    \node [circle, fill=celestial_blue!72, right of=v5, node distance=3cm, minimum width=1.8em, align=center, inner sep=0, outer sep=0, font=\scriptsize] (v_t) {$v_{\scaleto{t+1}{2.8pt}}$};
    
    \node [emb_layer, above of=v_t, node distance=1cm, minimum width=6em] (t_emb) {embedding layer};
    
     \node[v_rep, rectangle split part fill={celestial_blue!48}, above of=t_emb, node distance=2.5cm] (t_r) {
      \nodepart{four} $\,\cdots$
     };
     
     \node [right of=v_i, node distance=1.5cm] (v_s) {$\bm{s}$};
     \node [left of=t_r, node distance=1.6cm] (t_o) {$\bm{v}_{t+1}$};
    
    \node [left of=v1, node distance=4cm] (b_dots) {$\cdots$};
     \node [aug2, left of=b_dots, node distance=0.6cm] (b2) {$\widetilde{v}_2$};
    \node [aug2, left of=b2, node distance=0.75cm] (b1) {$\widetilde{v}_1$};
    \node [aug2, right of=b_dots, node distance=0.6cm, inner sep=0, outer sep=0] (b4) {$\widetilde{v}_{\scaleto{t-1}{2.8pt}}$};
    \node [aug2, right of=b4, node distance=0.75cm] (b5) {$\widetilde{v}_t$};
    \draw [thick,draw=black!42, decorate,decoration={brace,amplitude=8pt,mirror}] ([xshift=-1mm, yshift=-1mm] b1.south west) -- ([xshift=1mm, yshift=-1mm] b5.south east) node[midway,yshift=-0.5cm] (b_a) {$s^{a_j}$};
    
    \node [emb_layer, above of=b_dots, node distance=1cm] (b_emb) {embedding layer};
    \node [encoder, above of=b_emb, node distance=1.2cm] (b_enc) {encoder~ $f(\cdot)$};
    \node [dashed, draw=black, fit={([yshift=-0.5mm, xshift=-0.5mm] b_emb.south west) ([yshift=0.5mm, xshift=0.5mm] b_enc.north east)}, inner sep=3] (b_m) {};
    
    \node[v_rep, rectangle split part fill={casablanca!62}, above of=b_enc, node distance=1.3cm] (v_b) {
    \nodepart{four} $\,\cdots$
     };
      \node [left of=v_b, node distance=1.6cm] (v_ba) {$\bm{s}^{a_j}$};
    
    \node [left of=b1, node distance=4cm] (a1_dots) {$\cdots$};
    \node [aug1, left of=a1_dots, node distance=0.6cm] (a2) {$\widehat{v}_2$};
    \node [aug1, left of=a2, node distance=0.75cm] (a1) {$\widehat{v}_1$};
    \node [aug1, right of=a1_dots, node distance=0.6cm] (a4) {$\widehat{v}_{\scaleto{t-1}{2.8pt}}$};
    \node [aug1, right of=a4, node distance=0.75cm] (a5) {$\widehat{v}_t$};
    \draw [thick,draw=black!42, decorate,decoration={brace,amplitude=8pt,mirror}] ([xshift=-1mm, yshift=-1mm] a1.south west) -- ([xshift=1mm, yshift=-1mm] a5.south east) node[midway,yshift=-0.5cm] (a_a) {$s^{a_i}$};

    \node [emb_layer, above of=a1_dots, node distance=1cm] (a_emb) {embedding layer};
    \node [encoder, above of=a_emb, node distance=1.2cm] (a_enc) {encoder~ $f(\cdot)$};
    \node [dashed, draw=black, fit={([yshift=-0.5mm, xshift=-0.5mm]a_emb.south west) ([yshift=0.5mm, xshift=0.5mm] a_enc.north east)}, inner sep=3] (a_m) {};
    
    \node[v_rep, rectangle split part fill={flamingo!62}, above of=a_enc, node distance=1.3cm] (v_a) {
    \nodepart{four} $\,\cdots$
     };
     
     \node [right of=v_a, node distance=1.6cm] (v_aa) {$\bm{s}^{a_i}$};

     \draw[DFARROW, dashed] (a_m) -- (b_m) node [pos=0.5, above, black] (ab){share};
     \draw[DFARROW, dashed] (b_m) -- (o_m) node [pos=0.5, above, black] {share};

     \node [above of=v_a, node distance=1cm, xshift=26.75mm] (cl_score) {$\bm{s}^{a_i} \cdot \bm{s}^{a_j}$};
     \node [above of=cl_score, node distance=0.8cm] (cl_loss) {contrastive loss};
     \draw[FARROW] (cl_score) -> (cl_loss);
     \draw[FARROW, rounded corners,] (v_a.north) |- (cl_score.west) {} ;
     \draw[FARROW, rounded corners,] (v_b.north) |- (cl_score.east) {} ;
     
     \node [above of=v_i, node distance=1cm, xshift=21.75mm] (rec_score) {$\bm{s} \cdot \bm{v}_{t+1}$};
     \node [above of=rec_score, node distance=0.8cm] (rec_loss) {recommendation loss};
     \draw[FARROW] (rec_score) -> (rec_loss);
     \draw[FARROW, rounded corners,] (v_i.north) |- (rec_score.west) {} ;
     \draw[FARROW, rounded corners,] (t_r.north) |- (rec_score.east) {} ;
    
     \foreach \x in {1,2,4,5}
     {
        \draw[FARROW] (a\x) -- ++(0, 0.7);
        \draw[FARROW] (b\x) -- ++(0, 0.7);
        \draw[FARROW] (v\x) -- ++(0, 0.7);
        \draw[FARROW] ($(a\x.north)+(0, 1.03)$) -- ++(0, 0.4);
        \draw[FARROW] ($(b\x.north)+(0, 1.03)$) -- ++(0, 0.4);
        \draw[FARROW] ($(v\x.north)+(0, 1.03)$) -- ++(0, 0.4);
     }
     \draw[FARROW] (v_t) -> (t_emb);

    \draw[FARROW] (a_enc) -> (v_a);
    \draw[FARROW] (b_enc) -> (v_b);
    \draw[FARROW] (o_enc) -> (v_i);
    \draw[FARROW] (t_emb) -> (t_r);
    \draw[FARROW, dashed] (o_a.south) -- ++(0, -0.3) -- ++(-5.3, 0) -> ++(0, 0.4);
    \draw[FARROW, dashed] (o_a.south) -- ++(0, -0.3) -- ++(-10.7, 0) -> ++(0, 0.4);
    
    \node [right of=a_a, node distance=1.7cm, yshift=-3mm] (aug_1) {$a_i \sim \mathcal{A}$};
    \node [right of=b_a, node distance=1.7cm, yshift=-3mm] (aug_1) {$a_j \sim \mathcal{A}$};

    \end{tikzpicture} }
    \caption{A simple framework for CL4SRec.
    Two data augmentation methods, $a_i$ and $a_j$, are sampled from the same augmentation set $\mathcal{A}$. 
    They are applied to each user's sequence and then we can obtain two correlated views of each sequence.
    A shared embedding layer and the user representation model $f(\cdot)$ transform the original and augmented sequences to the latent space where the contrastive loss and recommendation loss are applied.}
    \label{fig:multitask}
\end{figure*}

The discrimination tasks adopt a contrastive learning framework to make a comparison between different samples.
Previous works have already achieved promising results in CV area~\cite{Simclr,Memory_bank,Momentum_contrast}.
Even without supervised signals during pre-training, contrastive learning methods can outperform strong baselines on various CV tasks~\cite{Simclr}.
To make a more difficult comparison, \citet{Memory_bank} designed a shared memory bank to increase the batch size during training, thus enhancing the final performance.
Later, \citet{Momentum_contrast} improved that with a dynamic queue and a moving-averaged encoder.

When it comes to recommendation systems, existing works apply self-supervised learning to improve recommendation performance.
\citet{Google} presented a self-supervised learning framework to learn item representations combined with a set of categorical features.
\citet{S3} proposed to learn the correlations among attributes, items, and sequences via auxiliary self-supervised objectives.
Unlike these methods, this work focuses on exploring contrastive learning on user behavior sequence to improve user representation learning.

\section{CL4SRec}
In this section, we present our contrastive learning framework for sequential recommendation~(CL4SRec), which only utilizes information of users' historical behaviors. 
We first introduce the notations used in this paper and formulate the sequential recommendation problem in Section 3.1.
Then, a general contrastive learning framework is introduced in Section 3.2.
In Section 3.3, we propose three augmentation methods to construct the contrastive tasks.
In section 3.4, we introduce the user representation model used in our approach.
Since our CL4SRec is a general framework, we select one of the state-of-the-art models---the encoder of Transformer~\cite{Transformer}
as our user representation model.
Finally, we propose how to train the user representation model via a multi-task learning framework.

\subsection{Notations and Problem Statement}
In this paper, we represent column vectors and matrices by bold italic lower case letters (e.g., $\bm{u}$, $\bm{v}$) and bold upper case letters (e.g., $\bm{R}$), respectively.
The $j$-th row of a matrix $\bm{R}$ is represented by $\bm{R}_j^{\top}$.
And we use calligraphic letters to represent sets (e.g., $\mathcal{U}$, $\mathcal{V}$, $\mathcal{A}$). 

Let $\mathcal{U}$ and $\mathcal{V}$ denote a set of users and items, respectively, where $|\mathcal{U}|$ and $|\mathcal{V}|$ denote the numbers of users or items. 
We represent a user or an item with  $u \in \mathcal{U}$ or $v \in \mathcal{V}$.
In sequential recommendation tasks, users' behavior sequences are usually in a chronological order.
Therefore, we represent the interaction sequence for user $u$ with $s_u=[v_1^{(u)},v_2^{(u)},\dots,v^{(u)}_{|s_u|}]$, where $v_t^{(u)}$ denotes the item which user $u$ interacts at the time %
step $t$ and $|s_u|$ denotes the length of interaction sequence for user $u$.
And $s_{u,t}=[v_1^{(u)},v_2^{(u)},\dots,v^{(u)}_{t}]$ denotes the subsequence of user $u$, which only focus on items interacted before $t+1$.
Also, let $\mathcal{A}$ denotes a set of augmentations that may be applied in the contrastive learning tasks.

Based on the above notations, we now define the task for the sequential recommendation. It focuses on predicting the most possible item which the user $u$ will interact with at the timestamp $|s_u| + 1$, given 
her/his historical interaction sequences without any other auxiliary contextual information.
It can be formulated as follows:
\begin{equation}
    v_u^* = \arg \max_{v_i \in \mathcal{V}} P\Bigl(v_{|s_u| + 1}^{(u)}=v_i \mid s_u\Bigr).
\end{equation}

\subsection{Contrastive Learning Framework}
Inspired by the SimCLR framework~\cite{Simclr} for learning visual representation,
we seek to explore applying contrastive learning algorithms to sequential recommendation task to obtain a powerful user representation model.  
The framework comprises three major components, including a stochastic data augmentation module, a user representation encoder, and a contrastive loss function. The framework is illustrated on the left side of Figure~\ref{fig:multitask}.
\tikzset{
  FARROW/.style={arrows={-{Latex[length=1.25mm, width=1.mm]}}, thick},
  origin/.style = {circle, fill=flamingo!72, minimum width=1.5em, align=center, inner sep=0, outer sep=0, font=\scriptsize},
  aug1/.style = {circle, fill=monte_carlo!72, minimum width=1.5em, align=center, inner sep=0, outer sep=0, font=\scriptsize},
  aug2/.style = {circle, fill=casablanca!72, minimum width=1.8em, align=center, inner sep=0, outer sep=0, font=\scriptsize},
}

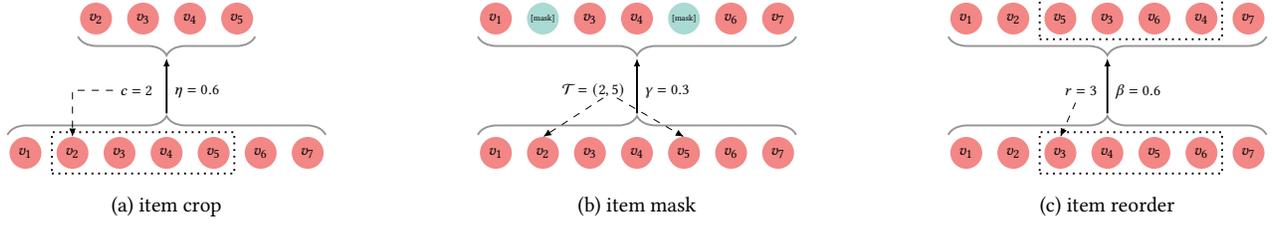
\begin{figure*}
\resizebox{0.95\linewidth}{!}{
    \begin{tikzpicture}
        
    \node[] (o) at (0, 0) {(c) item reorder};
    
    \node [origin, above of=o, node distance=0.8cm, align=center] (v4) {$v_4$};
    \node [origin, left of=v4, node distance=0.7cm] (v3) {$v_3$};
    \node [origin, left of=v3, node distance=0.7cm] (v2) {$v_2$};
    \node [origin, left of=v2, node distance=0.7cm] (v1) {$v_1$};
    \node [origin, right of=v4, node distance=0.7cm, inner sep=0, outer sep=0] (v5) {$v_5$};
    \node [origin, right of=v5, node distance=0.7cm] (v6) {$v_6$};
    \node [origin, right of=v6, node distance=0.7cm] (v7) {$v_7$};
    
    \node [thick, dotted, draw=black, fit={(v3) (v6)}, inner sep=2] (span) {};

    \foreach \x in {1,2,7}
    {
        \node [origin, above of=v\x, node distance=2cm] (a\x) {$v_{\x}$};
    } 
    \node [origin, above of=v3, node distance=2cm] (a3) {$v_5$};
    \node [origin, above of=v4, node distance=2cm] (a4) {$v_3$};
    \node [origin, above of=v5, node distance=2cm] (a5) {$v_6$};
    \node [origin, above of=v6, node distance=2cm] (a6) {$v_4$};
    
    \node [thick, dotted, draw=black, fit={(a3) (a6)}, inner sep=2] (span1) {};
    
    \draw [thick,draw=black!42, decorate,decoration={brace,amplitude=8pt}] ([xshift=-1mm, yshift=1mm] v1.north west) -- ([xshift=1mm, yshift=1mm] v7.north east) node[midway,yshift=0.5cm] (a_a) {};
    \draw [thick,draw=black!42, decorate,decoration={brace,amplitude=8pt, mirror}] ([xshift=-1mm, yshift=-1mm] a1.south west) -- ([xshift=1mm, yshift=-1mm] a7.south east) node[midway,yshift=0.5cm] (a_b) {};
    
     \draw[FARROW,] ([yshift=-3mm] a_a.north) -> ([yshift=-7mm] a_b.south) node[pos=0.4, right, align=left] (r_reorder) {\scriptsize $\beta = 0.6$} ;
     \node [left of=r_reorder, node distance=0.85cm, align=center, yshift=0.4pt] (ri) {\scriptsize $r=3$};
     \draw[FARROW, thin, dashed] (ri) -> (v3.north)  {} ;

    \node [left of=o, node distance=7cm, align=center] (mo) {(b) item mask};
    
    \foreach \x in {1,2,3,4,5,6,7}
    {
        \node [origin, left of=v\x, node distance=7cm] (mv\x) {$v_{\x}$};
    } 
    
    \foreach \x in {1,3,4,6,7}
    {
        \node [origin, above of=mv\x, node distance=2cm] (mva\x) {$v_{\x}$};
    } 
    \foreach \x in {2,5}
    {
     \node [aug1, above of=mv\x, node distance=2cm] (mva\x) { \scalebox{.6}{[mask]}};
     }
     \draw [thick,draw=black!42, decorate,decoration={brace,amplitude=8pt}] ([xshift=-1mm, yshift=1mm] mv1.north west) -- ([xshift=1mm, yshift=1mm] mv7.north east) node[midway,yshift=0.5cm] (mv_a) {};
    \draw [thick,draw=black!42, decorate,decoration={brace,amplitude=8pt, mirror}] ([xshift=-1mm, yshift=-1mm] mva1.south west) -- ([xshift=1mm, yshift=-1mm] mva7.south east) node[midway,yshift=0.5cm] (mv_b) {};
    \draw[FARROW,] ([yshift=-3mm] mv_a.north) -> ([yshift=-7mm] mv_b.south) node[pos=0.4, right, align=center] (r_mask) {\scriptsize $\gamma = 0.3$} ;
     \node [left of=r_mask, node distance=1.1cm, align=center, yshift=0.5pt] (m_i) {\scriptsize $\mathcal{T} = (2,5)$};
     \draw[FARROW, thin, dashed] ([xshift=1.5mm, yshift=1mm] m_i.south) -> (mv2.north)  {} ;
     \draw[FARROW, thin, dashed] ([xshift=3.5mm, yshift=1mm] m_i.south) -> (mv5.north) {} ;

    \node [left of=mo, node distance=7cm, align=center] (co) {(a) item crop};
    
    \foreach \x in {1,2,3,4,5,6,7}
    {
        \node [origin, left of=mv\x, node distance=7cm] (cv\x) {$v_{\x}$};
    } 
    
    \foreach \x in {2,3,4,5}
    {
        \node [origin, above of=cv\x, node distance=2cm, xshift=0.35cm] (cva\x) {$v_{\x}$};
    } 
    
    \node [thick, dotted, draw=black, fit={(cv2) (cv5)}, inner sep=2] (cv_span1) {};
    \draw [thick,draw=black!42, decorate,decoration={brace,amplitude=8pt}] ([xshift=-1mm, yshift=1mm] cv1.north west) -- ([xshift=1mm, yshift=1mm] cv7.north east) node[midway,yshift=0.5cm] (cv_a) {};
    \draw [thick,draw=black!42, decorate,decoration={brace,amplitude=8pt, mirror}] ([xshift=-1mm, yshift=-1mm] cva2.south west) -- ([xshift=1mm, yshift=-1mm] cva5.south east) node[midway,yshift=0.5cm] (cv_b) {};
    
    \draw[FARROW,] ([yshift=-3mm] cv_a.north) -> ([yshift=-7mm] cv_b.south) node[pos=0.4, right, align=center] (r_crop) {\scriptsize $\eta = 0.6$} ;
    \node [left of=r_crop, node distance=0.9cm, align=center, yshift=0.4pt] (co) {\scriptsize $c=2$};
    \draw[FARROW, thin, dashed] (co) -| (cv2.north)  (r_crop) {} ;

    \end{tikzpicture}
    }
    \caption{A brief illustration of augmentation operations applied in our CL4SRec model, including item crop, item mask, and item reorder.}
    \label{fig:aug}
\end{figure*}

\subsubsection{Data Augmentation Module.} A stochastic data augmentation module is employed to transform each data sample randomly into two correlated instances.
If the two transformed instances are from the same sample, they are treated as the positive pair.
If they are transformed from different samples, they are treated as the negative pair.
In the sequential recommendation task, we apply two randomly sampled augmentation methods ($a_i \in \mathcal{A}$ and $a_j \in \mathcal{A}$) to each user's historical behaviors sequence $s_u$, and obtain two views of the sequence, denoting $s_u^{a_i}$ and $s_u^{a_j}$.

\subsubsection{User Representation Encoder.}
We utilize a neural network as a user representation encoder to extract information from the augmented sequences.
With this encoder, we can obtain meaningful user representations from their augmented sequences, which is $\bm{s}_u^a=f(s_u^a)$.
Since our CL4SRec has no constraints on the choice of user representation model, in this work, we adopt the Transformer %
encoder~\cite{Transformer} 
to encode the user representation, which has shown promising results in recent works~\cite{SASRec,Bert4rec}.
It is worth mentioning that 
SIMCLR also applies an auxiliary non-linear projection module after $f(\cdot)$.
However, we find that removing this auxiliary projection leads to a significant improvement in our experiments.
In this paper, we discard this additional projection component.

\subsubsection{Contrastive Loss Function.}
Finally, a contrastive loss function is applied to distinguish whether the two representations are derived from the same user historical sequence.
To achieve this target, 
the contrastive loss learns to minimize the difference between differently augmented views of the same user historical sequence and maximize the difference between the augmented sequences derived from different users.
Considering a mini-batch of $N$ users $u_1, u_2,\dots, u_N$, we apply two random augmentation operators to each user's sequence and obtain 2$N$ augmented sequences $[s_{u_1}^{a_i}, s_{u_1}^{a_j}, s_{u_2}^{a_i}, s_{u_2}^{a_j}, \dots, s_{u_N}^{a_i}, s_{u_N}^{a_j}]$. 
Similar to \citeauthor{Simclr}~\cite{Simclr} and \citeauthor{Google}~\cite{Google}, for each user $u$, we treat $(s_{u}^{a_i}, s_{u}^{a_j})$ as the positive pair, and treat other $2(N-1)$ augmented examples within the same minibatch as negative samples $S^{-}$. %
We utilize dot product to measure the similarity %
between each representation, $\mathrm{sim}(\bm{u}, \bm{v}) = \bm{u}^{\top}\bm{v}$.
Then the loss function~$\mathcal{L}_\mathrm{cl}$ for a positive pair $(s_{u}^{a_i}, s_{u}^{a_j})$ can be defined similarly to the widely used softmax cross entropy loss as:
\begin{equation}
    \!\mathcal{L}_{\mathrm{cl}}(s_{u}^{a_i}, s_{u}^{a_j} ){=}{-}\log\frac{\exp\bigl(\mathrm{sim}(\bm{s}_u^{a_i}, \bm{s}_u^{a_j})\bigr)}{\exp\Bigl(\!\mathrm{sim}(\bm{s}_u^{a_i}, \bm{s}_u^{a_j})\!\Bigr) {+} \scaleto{{\displaystyle \sum_{s^{-} {\in} S^{-}}}}{16pt}\!\! \exp\Bigl(\!\mathrm{sim}(\bm{s}_u^{a_i}, \bm{s}^{-})\!\Bigr)}. 
\end{equation}

\subsection{Data Augmentation Operators}
Based on the above contrastive learning framework, 
we next discuss the design of the transformations in the contrastive learning framework, which can incorporate additional self-supervised signals to enhance the user representation model. 
As shown in Figure~\ref{fig:aug}, we introduce three basic augmentation approaches that can construct different views of the same sequence but still maintain the main preference hidden in historical behaviors.

\subsubsection{Item Crop}
Random crop is a common data augmentation technique to increase the variety of images in computer vision.
It usually creates a random subset of an original image to help the model generalize better.
Inspired by the random crop technique in images, we propose the item crop augmentation method for the contrastive learning task in the sequential recommendation.
For each user's historical sequence $s_u$, we randomly select a continuous sub-sequence: $s_u^{\scaleto{\mathtt{crop}}{3pt}}=[v_{c}, v_{c+1}, \dots, v_{\scaleto{c + L_{\mathtt{c}} {-} 1}{4.5pt}}]$ with the sequence length $L_{\mathtt{c}} = \lfloor \eta *|s_u| \rfloor $.
Here, we omit the superscript $u$ for each item $v$ for simplicity.
This augmentation method can be formulated as:
\begin{equation}
   s_u^{\scaleto{\mathtt{crop}}{3pt}} = a_{\scaleto{\mathtt{crop}}{3pt}}(s_u) = \left[v_{c}, v_{c+1}, \dots, v_{\scaleto{c + L_{\mathtt{c}} {-} 1}{4.5pt}}\right].
\end{equation}

The effect of our item crop augmentation method can be explained in two aspects.
Firstly, it provides a local view of the user's historical sequence.
It enhances the user representation model by learning a generalized user preference without comprehensive information of users.
Secondly, under the contrastive learning framework, if the two cropped sequences have no intersection, it can be regarded as a next sentence prediction task~\cite{BERT}.
It pushes the model to predict the change of the user's preference.

\subsubsection{Item Mask}
The technique of randomly zero-masking input word, which is also called ``word dropout'', is widely adopted to avoid overfittings in many natural language processing tasks, such as sentence generation~\cite{WordDropout1},  sentiment analysis~\cite{worddroput2}, and question answering~\cite{worddropout3}.
Inspired by this word dropout technique, we propose to apply a random item mask as one of the augmentation methods for contrastive learning.
For each user historical sequence $s_u$, we randomly mask a proportion $\gamma$ of items $\mathcal{T}_{s_u}=(t_1, t_2, \dots, t_{L_{\mathtt{m}}})$ with the length $L_{\mathtt{m}} = \left\lfloor \gamma * |s_u| \right\rfloor$.
Here, $t_{j}$ is the index of items in $s_{u}$ which will be masked.
If the item in the sequence is masked, it is replaced by a special item $[\mathrm{mask}]$.
Therefore, this augmentation method can be formulated as:
\begin{equation}
\begin{aligned}
    s_u^{\scaleto{\mathtt{mask}}{3pt}} &= a_{\scaleto{\mathtt{mask}}{3pt}}(s_u) = \left[\widehat{v}_1, \widehat{v}_2,\dots,\widehat{v}_{|s_u|}\right], \\
   \widehat{v}_t &= \left\{ \begin{aligned} 
 &\, v_t, & t \notin \mathcal{T}_{s_u} \\ 
 &\, [\mathrm{mask}], & t \in \mathcal{T}_{s_u}. \\ 
\end{aligned} 
\right.
\end{aligned}
\end{equation}

Since a user's intention is relatively stable over a period of time, the user's historical interacted items mostly reflect a similar purpose.
For example, if a user intends to purchase a pair of sports shoes, (s)he may click many sports shoes to decide which ones to buy.
Thus, with our item mask augmentation, the two different views derived from the same user sequence can still reserve the main intention of the user.
In this way, this self-supervised signal can prevent the user representation encoder from co-adapting too much.

\subsubsection{Item Reorder}
Many approaches employ the strict order assumption that most adjacent items in the users' historical sequences are sequentially dependent~\cite{FPMC,GRU4Rec}.
However, in the real world, sometimes the order of users' interactions are in a flexible manner~\cite{SeqSurvey,caser}, due to various unobservable external factors~\cite{YoutubeDNN}. 
In such a situation, we can derive a self-supervised augmentation operator to capture the sequential dependencies under the assumption of flexible order.
With this operator, we can encourage the user representation model to rely less on the order of interaction sequences, thus enhancing the model to be more robust when encountering new interactions. 

For this purpose, we adopt the item reorder task as another augmentation method for contrastive learning.
Inspired by swap operations in the natural language processing~\cite{Swap1, Swap2}, we alter the order of items in the user sequence with a proportion of $\beta$ by randomly shuffling their positions. 
More concretely, for each user historical sequence $s_u$, we randomly shuffle a continuous  sub-sequence $ [v_r, v_{r+1}, \dots, v_{\scaleto{r + L_{\mathtt{r}} {-} 1}{4.5pt}}]$, which starts at $r$ with length $L_{\mathtt{r}} = \left\lfloor \beta * |s_u| \right\rfloor$, to $ [\widehat{v}_r, \widehat{v}_{r+1}, \dots, \widehat{v}_{\scaleto{r + L_{\mathtt{r}} {-} 1}{4.5pt}}]$.
This augmentation method can be formulated as:
\begin{equation}
    s_u^{\scaleto{\mathtt{reorder}}{3pt}} = a_{\scaleto{\mathtt{reorder}}{3pt}}(s_u) = \left[v_1,v_2,\dots, \widehat{v}_{i},\dots,\widehat{v}_{\scaleto{i + L_{\mathtt{r}} {-} 1}{4.5pt}}, \dots, v_{|s_u|}\right].
\end{equation}

\subsection{User Representation Model}
\tikzset{
  emb/.style = {draw, rectangle, fill=flamingo!62, minimum width=3em, minimum height=1.5em, inner sep=0, outer sep=0},
  transformer/.style = {draw, rectangle, fill=Madang, minimum width=3em, minimum height=1.5em, font=\scriptsize},
  norm/.style = {draw, rectangle, fill=spring_sun, minimum width=4.5em, minimum height=1em, inner sep=0, outer sep=0},
  dropout/.style = {draw, rectangle, fill=hint_green, minimum width=4.5em, minimum height=1em, inner sep=0, outer sep=0},
  mh/.style = {draw, rectangle, fill=dairy_cream, minimum width=4.5em, minimum height=1.5em},
  ff/.style = {draw, rectangle, fill=french_pass, minimum width=4.5em, minimum height=1.5em},
  posemb/.style = {draw, rectangle, fill=watusi, minimum width=3em, minimum height=1.3em, inner sep=0, outer sep=0},
  proj/.style = {draw, rectangle, fill=Madang, minimum width=3em, minimum height=1.5em, inner sep=0, outer sep=0},
  FARROW/.style={arrows={-{Latex[length=1mm, width=0.8mm]}}}
}

\begin{figure}
\centering
\resizebox{0.46\textwidth}{!}{
\begin{tikzpicture}

\node[] (e1) at (0, 0) {};
\node[right of=e1, node distance=1.2cm] (e2) {};
\node[right of=e2, node distance=1.2cm] (e3) {};
\node[right of=e3, node distance=2cm] (e4) {};

\node[emb, above of=e1, node distance=0.6cm, minimum height=1.2em] (sa1) {$\bm{v}_1$};
\node[emb, above of=e3, node distance=0.6cm, minimum height=1.2em] (sa3) {$\bm{v}_{\scaleto{|s_u|-1}{5pt}}$};
\node[emb, above of=e4, node distance=0.6cm, minimum height=1.2em] (sa4) {$\bm{v}_{\scaleto{|s_u|}{5pt}}$};
\foreach \x in {1,3,4}
{
\node[transformer, above of=sa\x, node distance=1cm, minimum height=1.2em, inner sep=0, outer sep=0] (st\x) {Trm};
\node[transformer, above of=st\x, node distance=1cm, minimum height=1.2em, inner sep=0, outer sep=0] (stt\x) {Trm};
}

\node[right of=sa1, node distance=1.2cm, minimum height=1.2em] (sa2) {$\ldots$};

\node[posemb, below of=sa1, node distance=0.8cm] (pe1) {$\bm{p}_1$};
\node[below of=sa2, node distance=0.8cm] (pe2) {$\cdots$};
\node[posemb, below of=sa3, node distance=0.8cm] (pe3) {$\bm{p}_{\scaleto{|s_u|-1}{5pt}}$};
\node[posemb, below of=sa4, node distance=0.8cm] (pe4) {$\bm{p}_{\scaleto{|s_u|}{5pt}}$};

\node[below of=pe1, node distance=0.7cm] (si1) {$v_1$};
\node[below of=pe3, node distance=0.7cm] (si3) {$v_{\scaleto{|s_u|-1}{5pt}}$};
\node[below of=pe4, node distance=0.7cm] (si4) {$v_{\scaleto{|s_u|}{5pt}}$};

\draw[FARROW] (si1.east) -- ++(0.45, 0) |-  (sa1.east) ;
\foreach \x in {1,3,4}
{
\node[below of=sa\x, node distance=0.4cm] (add\x) {\large\textbf{+}};
\draw[FARROW] (si\x.east) -- ++(0.45, 0) |-  (sa\x.east) ;
\draw[FARROW] (si\x) -- (pe\x) ;
}

\node[below of=pe2, node distance=0.7cm, minimum height=1.2em] (si2) {$\ldots$};
\node[above of=sa2, node distance=0.8cm, minimum height=1.2em] (st2) {$\ldots$};
\node[above of=st2, node distance=0.8cm, minimum height=1.2em] (stt2) {$\ldots$};
\node[above of=stt1, node distance=0.7cm] (so1) {$v_2$};
\node[above of=stt3, node distance=0.7cm] (so3) {$v_{\scaleto{|s_u|}{5pt}}$};
\node[above of=stt4, node distance=0.7cm] (so4) {$v_{\scaleto{|s_u|+1}{5pt}}$};

\foreach \x in {1,3,4}
{
\draw[FARROW] (sa\x.north) -> (st\x.south) ;
\draw[FARROW] (st\x.north) -> (stt\x.south) ;
\draw[FARROW] (stt\x) -> ($(so\x.south)-(0,-0.08)$) ;
}

\foreach \x/\y in {1/3, 1/4, 3/4}
{
\draw[FARROW] (sa\x.north) -> (st\y.south) ;
\draw[FARROW] (st\x.north) -> (stt\y.south) ;
}

\draw[thick,dotted]  ($(si1.south west)+(-0.4, 0)$) rectangle ($(so4.north east)+(0.4, 0)$);
\node[below of=si2, node distance=1.1cm, xshift=1cm] (sas) {(a) SASRec model architecture.};

\node[mh, right of=e4, node distance=2.7cm, align=center, font=\scriptsize, yshift=-0.3cm] (trm1) {Multi-Head\\ Attention};
\node[dropout, above of=trm1, node distance=0.75cm, align=center, font=\scriptsize] (do1) {Dropout};
\node[norm, above of=do1, node distance=0.6cm, align=center, font=\scriptsize] (n1) {Add \& Norm};

\draw[FARROW] (trm1) edge (do1);
\draw[FARROW] (do1) edge (n1);

\node[ff, above of=n1, node distance=1cm, align=center, font=\scriptsize] (ff1) {Position-wise\\Feed-Forward};
\node[dropout, above of=ff1, node distance=0.75cm, align=center, font=\scriptsize] (do2) {Dropout};
\node[norm, above of=do2, node distance=0.6cm, align=center, font=\scriptsize] (n2) {Add \& Norm};

\draw[FARROW] (n1) edge (ff1);
\draw[FARROW] (ff1) edge (do2);
\draw[FARROW] (do2) edge (n2);

\node[below of=trm1, node distance=1.2cm, align=center, font=\small] (input) {input};
\node[above of=n2, node distance=0.8cm, align=center, font=\scriptsize] (out) {};
\node[right of=out, node distance=0.7cm, align=center, font=\scriptsize, yshift=-0.35cm] (aa) {Trm};

\draw[FARROW] (n2) edge (out);
\draw[FARROW] (input) edge (trm1);

\draw[FARROW] ($(input.north)-(0,-0.3)$) -- ++(-1, 0) |- (n1.west) ;
\draw[FARROW] ($(n1.north)-(0,-0.3)$) -- ++(-1, 0) |- (n2.west) ;

\draw[thick,dotted, fill opacity=0.1, fill=bgc]  ($(input.south west)+(-0.8, 0.6)$) rectangle ($(out.north east)+(0.8, -0.6)$);

\node[below of=input, node distance=0.5cm] (trm) {(b) Transformer Layer.};

\end{tikzpicture}}
    \caption{A brief architecture of SASRec model and Transformer Encoder Layer.}
    \label{fig:model}
\end{figure}
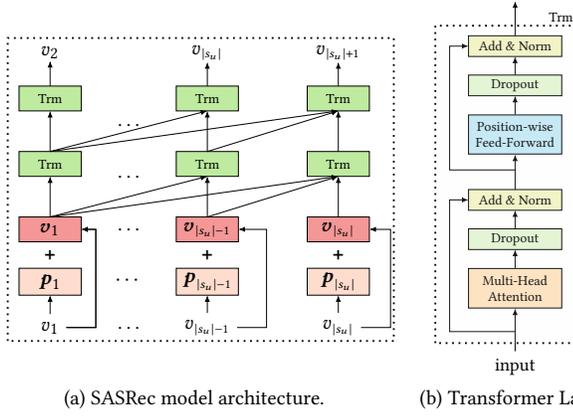
In this subsection, we describe how we model the user historical sequences by stacking the Transformer encoder.
We utilize the architecture of SASRec model in this paper, which applies a unidirectional Transformer encoder and has achieved promising results in the sequential recommendation task~\cite{SASRec}.
The Transformer encoder consists of three sub-layers: an embedding Layer, a multi-head self-attention module, and a position-wise feed-forward Network, which is briefly illustrated in Figure~\ref{fig:model}.

\subsubsection{The Embedding Layer}
The Transformer encoder utilizes an item embedding matrix $\bm{E} \in \mathbb{R}^{|\mathcal{V}|\times d}$ to project high dimensional one-hot item representations to low dimensional dense vectors.
In addition, to represent the position information of sequence, it leverages the learnable position embedding $\bm{P} \in \mathbb{R}^{T \times d}$ to capture this feature of sequences.
Note that the number of position vectors $T$ restricts the maximum length of the user's historical sequence.
Therefore, when we infer user $u$ representation at time step $t + 1$, 
, we truncate the input sequence $s_{u,t}$ to the last $T$ items if $t > T$:
\begin{equation}
    s_{u,t}=\left[v_{\scriptscriptstyle t-T + 1},v_{\scriptscriptstyle t-T+2},\dots,v_{t}\right].
\end{equation}
Finally, we can obtain the input representations of items in the user sequence by adding the item embedding and position embedding together as:
\begin{equation}
    \bm{h}_i^{0} = \bm{v}_{i}+\bm{p}_{t},
\end{equation}
where $\bm{v}_i \in \bm{E}$ is the representation of item $v_i$ in the user sequence $s_{u,t}$. Here, we omit the superscript $u$ for convenience.

\subsubsection{Multi-Head Self-Attention Module}
After the embedding layer, the Transformer encoder introduces the self-attention module~\cite{Transformer} to capture the dependencies between each item pair in the sequence, which is effective in sequence modeling in many tasks.
Moreover, to extract the information from different subspaces at each position, here we adopt the multi-head self-attention instead of a single attention function.
It first utilizes different linear projections to project the input representations into $h$ subspaces.
Then it applies the self-attention mechanism to each head and derives the output representations by concatenating the intermediates and projecting it once again.
The computation is as follows:
\begin{equation}
\begin{aligned}
    \mathrm{MH}(\bm{H}^l) &=\mathrm{concat}(\mathrm{head}_1;\mathrm{head}_2;\cdots;\mathrm{head}_h)\bm{W}^O \\
    \mathrm{head}_i &= \mathrm{Attention}(\bm{H}^{l}\bm{W}_i^Q, \bm{H}^{l}\bm{W}_i^K, \bm{H}^{l}\bm{W}_i^V),
\end{aligned}
\end{equation}
where $\bm{W}_i^Q, \bm{W}_i^K, \bm{W}_i^V \in \mathbb{R}^{d \times \frac{d}{h}}$ and $\bm{W}^O \in \mathbb{R}^{d \times d}$.
Here, $\bm{H}^l$ is the input for the $(l + 1)$-th layer. 
The $\mathrm{Attention}$ operation is implemented by the scaled dot-product attention as follows:
\begin{equation}
    \mathrm{Attention}(\bm{Q}, \bm{K},\bm{V})
    = \mathrm{softmax}\left(\frac{\bm{Q}\bm{K}^{\top}}{\sqrt{d/h}}\right)\bm{V},
\end{equation}
where $\bm{Q}$, $\bm{K}$, and $\bm{V}$ represent the queries, the keys, and the values, respectively.
The factor $\sqrt{d/h}$ in this attention module is the scale factor to avoid large values of the inner product.

In sequential recommendation, we can only utilize the information before the time step $t$ when we predict the next item $v_{t+1}$.%
Theref\-ore, we apply the mask operations to the attention mechanism to discard the connections between $\bm{Q}_i$ and $\bm{K}_j$ where $j > i$.
These operations can avoid information leakage and are beneficial to the model training. 

\subsubsection{Position-wise Feed-Forward Network}
Though multi-head self-attention is beneficial to extract useful information from previous items, it is based on simple linear projections. 
We endow the model with nonlinearity with a position-wise feed-forward network.
It is applied at each position of the above sub-layer's output with shared learnable parameters:
\begin{equation}
\begin{aligned}
    \mathrm{PFFN}(\bm{H}^l)
    &= [\mathrm{FFN}(\bm{h}_1^l)^{\top}; \mathrm{FFN}(\bm{h}_2^l)^{\top}; \cdots;
    \mathrm{FFN}(\bm{h}_{|s_u|}^l)^{\top}] \\
    \mathrm{FFN}(\bm{h}_i^l)
    &= \mathrm{RELU}(\bm{h}_i^l\bm{W}_1 + \bm{b}_1)\bm{W}_2 + \bm{b}_2.
\end{aligned}
\end{equation}

\subsubsection{Stacking More Blocks}
Stacking more blocks is usually beneficial to learn more complex patterns for deep learning methods. 
However, with more parameters and a deeper network, the model becomes harder to converge.
To alleviate this problem, We employ several mechanisms, including the residual connection, layer normalization, and dropout module as follows:
\begin{equation}
    \mathrm{LayerNorm}(\bm{x} + \mathrm{Dropout}(\mathrm{sublayer}(\bm{x}))),
\end{equation}
where $\mathrm{sublayer(\cdot)}$ is the above multi-head self-attention operation and position-wise feed-forward network.
These mechanisms are widely adopted to stabilize and accelerate model training.

\subsubsection{User Representations}
Based on several Transformer blocks, we obtain the user representation at each time step $t$, which extracts useful information from the items interacted with before $t$.
Since our task is to predict the item at the time step $|s_u|+1$ for each user $u$, we set the final representation for user $u$ as her preference vector at time $|s_u|+1$:
\begin{equation}
\bm{s}_u = [\texttt{Trm}^L(s_u)]^{\top}_{|s_u|},
    \label{eq:sasrec}
\end{equation}
where $L$ is the number of stacking Transformer layers.
And the \texttt{Trm} function is the composition of following operations:
\begin{equation}
\begin{aligned}
    \texttt{Trm}(\bm{H}^{l}) &= \mathrm{LayerNorm}(\bm{F}^{l-1} + \mathrm{Dropout}(\mathrm{PFFN}(\bm{F}^{l-1})))\\
    \bm{F}^{l-1} &= \mathrm{LayerNorm}(\bm{H}^{l-1} + \mathrm{Dropout}(\mathrm{MH}(\bm{H}^{l-1}))).
\end{aligned}
\end{equation}

\subsection{Multi-task Training}
To leverage the self-supervised signals derived from the unlabeled raw data to enhance the performance of sequential recommendation, we adopt a multi-task strategy where the main sequence prediction task and the additional contrastive learning task are jointly optimized.
The total loss is a linear weighted sum as follows:
\begin{equation}
    \mathcal{L}_{\mathrm{total}} = \mathcal{L}_{\mathrm{main}} + \lambda\mathcal{L}_{\mathrm{cl}}.
\end{equation}
We adopt the negative log likelihood with sampled softmax as the main loss for each user $u$ at each time step $t + 1$ as: 

\begin{equation}
    \mathcal{L}_{\mathrm{main}}(s_{u,t} ){=}{-}\log\frac{\exp\bigl(\bm{s}_{u,t}^{\top}\bm{v}_{t+1}^{+}\bigr)}{\exp\bigl(\bm{s}_{u,t}^{\top}\bm{v}_{t+1}^{+}\bigr){+}\scaleto{{\displaystyle \sum_{\bm{v}_{t+1}^{-} {\in} \mathcal{V}^{-}}}}{16pt}\!\! \exp\Bigl(\!\bm{s}_{u,t}^{\top}\bm{v}_{t+1}^{-}\!\Bigr)}, 
\end{equation}
where $\bm{s}_{u,t}$, $\bm{v}_{t+1}^{+}$, and  $\bm{v}_{t+1}^{-}$ indicate the inferred user representation, the item which user $u$ interacts, and the randomly sampled negative item at the time step $t + 1$, respectively.

\section{Experiments}
In this section, we conduct extensive experiments to answer the following research questions:
\begin{itemize}
    \item[\textbf{RQ1.}] How does the proposed CL4SRec framework perform compared to the state-of-the-art baselines in the sequential recommendation task?
    \item[\textbf{RQ2.}] How do different augmentation methods impact the performance? 
    What is the influence of different augmentation hyper-parameters on CL4SRec performance?
    \item[\textbf{RQ3.}] How does the weight $\lambda$ of contrastive learning loss impact on the performance under the multi-task framework?
    \item[\textbf{RQ4.}] 
    How do different components of CL4SRec benefit its performance, i.e., augmentation methods and contrastive learning loss?
    \item[\textbf{RQ5.}]
    Does our CL4SRec really learn a better representation in the user behavior sequence level compared with other state-of-the-art baselines?
\end{itemize}

\subsection{Experiments Settings}
\subsubsection{Datasets}
We conduct experiments on four public datasets collected from the real-world platforms. 
Two of them are obtained from Amazon, one of the largest e-commercial platforms in the world. 
They have been introduced in \cite{Amazon}, which are split by top-level product categories in amazon.
In this work, we follow the settings in \cite{S3} and adopt two categories, ``Beauty'' and ``Sports and Outdoors''
Another dataset is collected by Yelp~\footnote{\url{https://www.yelp.com/dataset}}, 
which is a famous business recommendation platform for restaurants, bars, beauty salons, etc.
We follow the settings in \cite{S3} and use the transaction records after January 1st, 2019.
The last dataset is MovieLens-1M\footnote{\url{https://grouplens.org/datasets/movielens/1m/}}~(\textbf{ML-1M}) \cite{movielens} dataset, which is widely used for evaluating recommendation algorithms. 

For dataset preprocessing, we follow the common practice in~\cite{SASRec,S3}.
We convert all numeric ratings or presence of a review to ``1'' and others to ``0''. 
Then, for each user, we discard duplicated interactions and then sort their historical items by the interacted time step chronologically to obtain the user interacted sequence.
It is worth mentioning that to guarantee each user/item with enough interactions, we follow the preprocessing procedure in \cite{S3,FPMC}, which only keeps the ``5-core'' datasets.
We discard users and items with fewer than 5 interaction records iteratively.
The processed data statistics are summarized in Table~\ref{tab:dataset}.

\begin{table}
  \caption{Dataset statistics (after preprocessing).}
  \label{tab:dataset}
  \begin{tabular}{cccccc}
    \toprule
    Dataset &\#users &\#items &\#actions &avg.length & density\\
    \midrule
    Beauty & 22,363 & 12,101 & 198,502 & 8.8 & 0.07\%\\
    Sports & 25,598& 18,357 & 296,337 & 8.3 & 0.05\%\\
    Yelp & 30,983 & 29,227 & 321,087 & 10.3 & 0.04\%\\
    ML-1M & 6,040 & 3,953 & 1,000,209 & 165.6 & 4.19\%\\
  \bottomrule
\end{tabular}
\end{table}

\subsubsection{Evaluation}
We adopt the leave-one-out strategy to evaluate the performance of each method, which is widely employed in many related works~\cite{NCF,SASRec,S3}.
For each user, we hold out the last interacted item as the test data and utilize the item just before the last as the validation data.
The remaining items are used for training. 
To speed up the computation of metrics, many previous works use sampled metrics and only rank the relevant items with a smaller set of random items.
However, this sample operation may lead to inconsistent with their non-sampled version as illustrated by \citeauthor{Sampled_metrics}~\cite{Sampled_metrics}.
Therefore, we evaluate each method on the whole item set without sampling and rank all the items that the user has not interacted with by their similarity scores.
We employ Hit Ratio (HR) and Normalized Discounted Cumulative Gain (NDCG) to evaluate the performance of each method which are widely used in related works~\cite{FPMC,SASRec}.
HR focuses on the presence of the positive item, while NDCG further takes the rank position information into account. 
In this work, we report HR and NDCG with $k=5,10,20$.

\subsubsection{Baselines}
To verify the effectiveness of our method, we compare it with the
following representative baselines:
\begin{itemize}%
    \item \textbf{Pop}. It is a non-personalized approach which recommends the same items for each user.
These items are the most popular items which have the largest number of interactions in the whole item set.
\item \textbf{BPR-MF}~\cite{BPR}. It is one of the representative non-sequential baselines. 
It utilizes matrix factorization to model users and items with the pairwise Bayesian Personalized Ranking (BPR) loss.
\item \textbf{NCF}~\cite{NCF}.It employs a neural network architecture to model non-sequential user-item interactions instead of the inner product used by matrix factorization.
\item \textbf{GRU4Rec$^+$}~\cite{GRU4Rec,GRU4Rec+}. It applies GRU modules to model user sequences for session-based recommendation with ranking loss and is improved with a new class of loss functions and sampling strategy.
\item \textbf{SASRec}~\cite{SASRec}. It is one of the state-of-the-art baselines to solve the sequential recommendation task. It models user sequences through self-attention modules to capture users' dynamic interests.
\item \textbf{GC-SAN}~\cite{GCSAN}. It combines GNN with self-attention mechanism to capture both local and long-range transitions of neighbor items hidden in each interaction session.
\item \textbf{S$^3$-Rec$_{\mathrm{MIP}}$}~\cite{S3}.
It also utilizes the self-supervised learning methods to derive the intrinsic data correlation. However, it mainly focuses on how to fuse the context data and sequence data. In this section, we only compare the mask item prediction~(MIP) in S$^3$-Rec for fairness.

\end{itemize}

\subsubsection{Implementation Details}
We utilize the public implementation of BPR-MF, NCF, and GRU4Rec provided by \citet{Chorus}\footnote{\url{https://github.com/THUwangcy/ReChorus}} in PyTorch.
For other methods, we implement them by PyTorch as well.
For all models with learnable embedding layers, we set the embedding dimension size $d=64$, following previous works~\cite{SASRec,S3,Bert4rec}.
For each baseline, all other hyper-parameters are set following the suggestions from the original settings in their papers and we report each baseline performance under its optimal settings.

When it comes to our CL4SRec method, we initialize all parameters by the truncated normal distribution in the range $[-0.01, 0.01]$.
We use Adam optimizer~\cite{adam} to optimize the parameters with the learning rate of $0.001$, $\beta_1 = 0.9$, $\beta_2 = 0.999$, and linear decay of the learning rate, where the batch size is set to 256.
We train the model with early stopping techniques according to the performance on the validation set.
To investigate the effect of each two augmentation method combination, we fixed the augmentation methods used in CL4SRec each time.
And we test the crop/mask/reorder proportion of items from 0.1 to 0.9.
For our user representation model of CL4SRec, we stack 2 self-attention blocks together and set the head number as 2 for each block.
For a fair comparison, following the settings in~\cite{S3}, the maximum sequence length of $T$ is set to $20$ and $50$ for ML-1M and other datasets, respectively\footnote{We find that $T=20$ performs not worse than $T=50$ on the ML-1M dataset}.

\begin{table*}[]
\setlength{\tabcolsep}{0.7em}
    \centering
    \caption{Performance comparison of different methods on top-$N$ recommendation. Bold scores are the best in method group, while underlined scores are the second best. Improvements over baselines are statistically significant with $p < 0.01$.
    The reported result of CL4SRec for each dataset is the best result of applying one of the augmentations.
    } %
    \begin{adjustbox}{max width=\textwidth}
        \begin{tabular}{l l c c c c c c c c c}
        \toprule
        Datasets & Metric & Pop & BPR-MF & NCF & GRU4Rec$^+$  & SASRec & GC-SAN& S$^3$-Rec$_{\mathrm{MIP}}$ & CL4SRec &Improv.\\
        \midrule
        \multirow{6}{*}{Beauty} 
        & HR@5 & 0.0072 & 0.0120  & 0.0150  & 0.0239  & \underline{0.0347} &0.0270& 0.0327 & \textbf{0.0396} & 14.12\%\\
        
         & HR@10 & 0.0114 & 0.0299  & 0.0293 & 0.0399 & \underline{0.0630} &0.0444& 0.0566 & \textbf{0.0681} & ~~8.10\%\\
         
         & HR@20 & 0.0195 & 0.0524 & 0.0519 & 0.0637  & \underline{0.1007} &0.0698& 0.0905 & \textbf{0.1056} & ~~4.87\%\\
         
         & NDCG@5 & 0.0040 & 0.0065 & 0.0084 & 0.0150  & 0.0185 &0.0169&\underline{0.0193} & \textbf{0.0208} & ~~7.77\% \\
         
         & NDCG@10 & 0.0053 & 0.0122 & 0.0130 & 0.0201 & \underline{0.0276} &0.0225& 0.0270 & \textbf{0.0299} & ~~8.33\% \\
         
         & NDCG@20 & 0.0073  & 0.0179 & 0.0187 & 0.0261  & \underline{0.0371} &0.0289& 0.0355 & \textbf{0.0394} & ~~6.20\% \\
         \midrule
        \multirow{6}{*}{Sports} 
        & HR@5 & 0.0055 & 0.0092 & 0.0122 & 0.0155  & \underline{0.0185} &0.0174& 0.0171 & \textbf{0.0219} & 18.38\% \\
        
         & HR@10 & 0.0090 & 0.0188 & 0.0219 & 0.0259 &  \underline{0.0328} &0.0266& 0.0298 & \textbf{0.0387} & 17.98\% \\
         
         & HR@20 & 0.0149 & 0.0337 & 0.0373 & 0.0423 & \underline{0.0541} &0.0414& 0.0506 & \textbf{0.0595} & ~~9.98\% \\
         
         & NDCG@5 & 0.0040 & 0.0053 & 0.0069 & 0.0098  & 0.0101 &\underline{0.0110}& 0.0106 & \textbf{0.0116} & ~~5.45\%\\
         
         & NDCG@10 & 0.0051 & 0.0083 & 0.0100 & 0.0131 & \underline{0.0148} &0.0139& 0.0146 & \textbf{0.0171} & 15.54\%\\
         
         & NDCG@20 & 0.0066 & 0.0121  & 0.0139 & 0.0172 & \underline{0.0201} &0.0177& 0.0199 & \textbf{0.0228} &13.43\% \\
         \midrule
        \multirow{6}{*}{Yelp} & 
        HR@5 &  0.0056 & 0.0127 & 0.0131 & 0.0150 & 0.0157 & 0.0157&\underline{0.0186} & \textbf{0.0201} & ~~8.06\%\\
         
         & HR@10 & 0.0095 & 0.0273  & 0.0267 & 0.0276  &  0.0300 &0.0267& \underline{0.0313} & \textbf{0.0349} & ~~11.50\%\\
         
         & HR@20 & 0.0160 & 0.0500 & 0.0467 & 0.0484  & \underline{0.0529} & 0.0452&0.0519 & \textbf{0.0598} & 13.04\% \\
         
         & NDCG@5 & 0.0036 & 0.0074 & 0.0079 & 0.0091  & 0.0085 &0.0101&\underline{0.0116} & \textbf{0.0124} & ~~6.70\% \\
         & NDCG@10 & 0.0049 & 0.0121 & 0.0119 & 0.0131 & 0.0131 & 0.0136&\underline{0.0157} & \textbf{0.0171} & ~~8.92\%\\
         & NDCG@20 & 0.0065 &  0.0178 & 0.0172 & 0.0183 & 0.0188 & 0.0182&\underline{0.0209} & \textbf{0.0233} &11.48\%\\
         \midrule
        \multirow{6}{*}{ML-1M} 
        & HR@5 & 0.0078 & 0.0164 & 0.0164 & 0.0993 &  \underline{0.1108} & 0.1099 & 0.1078 & \textbf{0.1147} & ~~3.52\%\\
         
         & HR@10 & 0.0162 & 0.0354 & 0.0313 & 0.1806 &  0.1902 & 0.1906 & \underline{0.1952} & \textbf{0.1975} & ~~1.18\%\\
         
         & HR@20 & 0.0402 &  0.0712 & 0.0641 & 0.2892 &  \underline{0.3124} & 0.3104 &  0.3114 & \textbf{0.3174} & ~~1.60\%\\
         
         & NDCG@5 & 0.0052 & 0.0097  & 0.0096 & 0.0574 &  \underline{0.0648} & 0.0647 & 0.0616 & \textbf{0.0662} & ~~2.16\%\\
         
         & NDCG@10 & 0.0079 &  0.0158 & 0.0143 & 0.0835 & 0.0904 & 0.0895 & \underline{0.0917} & \textbf{0.0928} & ~~1.20\%\\
         
         & NDCG@20 & 0.0139 & 0.0248  & 0.0225 & 0.1108  & \underline{0.1211} & 0.1197 & 0.1204 & \textbf{0.1230} & ~~1.57\%\\
        \bottomrule
        \end{tabular}
    \end{adjustbox}
    \label{tab:result}
\end{table*}

\subsection{Overall Performance Comparison (RQ1)}
To answer \textbf{RQ1}, we compare the performance of our CL4SRec 
with the above baselines.
Table \ref{tab:result} summarizes the best results of all models on four datasets.
Note that the improvement columns are the performance of CL4SRec relative to the second best baselines.
Due to the space limitation, the results of CL4SRec reported in Table~\ref{tab:result} are the best results of using one of the three augmentations.

Based on the experiment results, we can observe that:
\begin{itemize}
\item The non-personalized method Pop exhibits the worst performance on all datasets since it ignores users' unique preferences hidden in their historical interactions.
Considering other baseline methods on all datasets, we observe that sequential methods (e.g., GRU4Rec$^+$ and SASRec) outperform non-sequential methods (e.g., BPR-MF and NCF) consistently. 
Compared with those non-sequential methods, these sequential methods utilize sequential information of users’ historical interactions, which contributes to improving performance in recommendation systems.
Among all the sequential models, SASRec achieves state-of-the-art performance on all datasets, which indicates that the powerful self-attention mechanism is suitable for capturing sequence patterns. 
 
\item When it comes to the GC-SAN, it does not achieve obvious improvements and even sometimes exhibits worse compared to the SASRec in our experimental settings.
One possible reason is that no rings are existing in each user sequence and the degree of each interaction is less than two in these datasets with the pre-processing precedure.
GNN cannot capture auxiliary useful information in this kind of sequences, but increases the burden of optimization.
For S$^3$-Rec$_{\mathrm{MIP}}$, it also sometimes exhibits worse compared to the SASRec.
Without auxiliary contextual information, it pre-trains and finetunes the base model on the exact same dataset, which may lead to catastrophic forgetting~\cite{catastrophic}.

\item Finally, according to the results, it is obvious that our proposed CL4SRec outperforms all baselines on all datasets in terms of all the evaluation metrics.
It achieves improvements on both sparse (e.g., Sports and Yelp) and dense datasets (e.g., ML-1m), especially on the sparse datasets.
CL4SRec gains 11.02\% on HR@5, 9.69\% on HR@10, 5.52\% on NDCG@5, and 8.50\% on NDCG@10 on average against the start-of-the-art baselines.
These experiments verify the effectiveness of our CL4SR\-ec method in the sequential recommendation task.
Different from S$^3$-Rec$_{\mathrm{MIP}}$ which still focus on sequence prediction issue, we adopt the contrastive learning framework to introduce other information, which enhances the user representation model to capture more accurate user representations.
\end{itemize}

\subsection{Comparison of Different Augmentation Methods (RQ2)}
To answer \textbf{RQ2}, we analyze how different augmentation operators and their proportion rates impact the performance.
To examine the effect of each augmentation operator, we only utilize one kind of operator in the contrastive learning task with the same proportion parameters each time.
Note that the high rate of $\gamma$ (mask) and $\beta$ (reorder) and low rate of $\eta$ (crop) in augmentations will make it more difficult to distinguish instances transformed from the same sequence from negative samples.
We report HR@20 and NDCG@20 as metrics for the experiments on the Sports and Yelp datasets due to the space limitation.

Figure~\ref{fig:Proportion2} demonstrates the performance of different augmentation methods with different proportion rates $\eta$, $\gamma$, and $\beta$ varing from 0.1 to 0.9.
We observe a few trends of different augmentation methods with the variation of proportion rates.
First, the performance of CL4SRec equipped with any augmentation method can outperform the SASRec baseline on all datasets for most choices of proportion rates.
It indicates the effectiveness of the proposed augmentation methods, since they all introduce auxiliary self-supervised signals hidden in raw data.
And we observe that none of the three augmentation operations can always achieve the best performance compared with other augmentations.
For example, Mask operation achieves the best results on the Sports dataset, but the Crop operation achieves the best results on the Yelp dataset.
This demonstrates that different augmentation methods are appropriate for different datasets since they focus on different aspects of the raw data.
\pgfplotsset{
axis background/.style={fill=gallery},
grid=both,
  xtick pos=left,
  ytick pos=left,
  tick style={
    major grid style={style=white,line width=1pt},
    minor grid style=bgc,
    draw=none
    },
  minor tick num=1,
  ymajorgrids,
	major grid style={draw=white},
	y axis line style={opacity=0},
	tickwidth=0pt,
}

\begin{figure}
\centering
    \begin{tikzpicture}[scale=0.47]
	\begin{groupplot}[
	    group style={group size=2 by 2,
	        horizontal sep = 50pt,
	        vertical sep=4em
	        }, 
	    width=0.5\textwidth,
	    xlabel=\large Proportion,
        ylabel=\large HR@20,
        xticklabels={0.1, 0.2, 0.3, 0.4, 0.5, 0.6, 0.7, 0.8, 0.9},
        xtick={1,2,3,4, 5, 6, 7, 8, 9},
        ymajorgrids,
        major grid style={draw=white},
        y axis line style={opacity=0},
        tickwidth=0pt,
        yticklabel style={
        /pgf/number format/fixed,
        /pgf/number format/precision=5
        },
        scaled y ticks=false,
        every axis title/.append style={at={(0.1,0.85)},font=\bfseries}
	    ]
		\nextgroupplot[
		legend style = {
		  font=\small,
          draw=none, 
          fill=none,
          column sep = 2pt, 
          /tikz/every even column/.append style={column sep=5mm},
          legend columns = -1, 
          legend to name = grouplegend2},
		title=Sports, 
		]
         \addplot[dashed, thick,color=free_speech_aquamarine,mark=triangle*] coordinates {
          (1, 0.0541)
          (2, 0.0541)
          (3, 0.0541)
          (4, 0.0541)
          (5, 0.0541)
          (6, 0.0541)
          (7, 0.0541)
          (8, 0.0541)
          (9, 0.0541)
        }; \addlegendentry{SASRec}
        \addplot[thick,color=matisse,mark=square] coordinates {
          (1, 0.0571)
          (2, 0.0596)
          (3, 0.0568)
          (4, 0.0582)
          (5, 0.0574)
          (6, 0.0550)
          (7, 0.0567)
          (8, 0.0545)
          (9, 0.0551)
        };\addlegendentry{Crop}
        \addplot[thick,color=sun_shade,mark=diamond] coordinates {
          (1, 0.0539)
          (2, 0.0540)
          (3, 0.0552)
          (4, 0.0556)
          (5, 0.0567)
          (6, 0.0563)
          (7, 0.0595)
          (8, 0.0585)
          (9, 0.0578)
        };
        \addlegendentry{Mask}
        \addplot[thick,color=flamingo,mark=*] coordinates {
          (1, 0.0508)
          (2, 0.0560)
          (3, 0.0545)
          (4, 0.0540)
          (5, 0.0553)
          (6, 0.0541)
          (7, 0.0550)
          (8, 0.0537)
          (9, 0.0541)
        };\addlegendentry{Reorder}
        
        \nextgroupplot[
        title=Sports,
        ylabel=\large NDCG@20]
         \addplot[dashed, thick,color=free_speech_aquamarine,mark=triangle*] coordinates {
          (1, 0.0201)
          (2, 0.0201)
          (3, 0.0201)
          (4, 0.0201)
          (5, 0.0201)
          (6, 0.0201)
          (7, 0.0201)
          (8, 0.0201)
          (9, 0.0201)
        };
        \addplot[thick,color=matisse,mark=square] coordinates {
          (1, 0.0216)
          (2, 0.0223)
          (3, 0.0220)
          (4, 0.0220)
          (5, 0.0217)
          (6, 0.0217)
          (7, 0.0219)
          (8, 0.0212)
          (9, 0.0211)
        };
        \addplot[thick,color=sun_shade,mark=diamond] coordinates {
          (1, 0.0214)
          (2, 0.0214)
          (3, 0.0210)
          (4, 0.0211)
          (5, 0.0215)
          (6, 0.0217)
          (7, 0.0228)
          (8, 0.0219)
          (9, 0.0219)
        };
        \addplot[thick,color=flamingo,mark=*] coordinates {
          (1, 0.0203)
          (2, 0.0213)
          (3, 0.0208)
          (4, 0.0207)
          (5, 0.0207)
          (6, 0.0209)
          (7, 0.0207)
          (8, 0.0206)
          (9, 0.0204)
        };

        \nextgroupplot[
        title=Yelp,
        ylabel=\large HR@20]
         \addplot[dashed, thick,color=free_speech_aquamarine,mark=triangle*] coordinates {
          (1, 0.0529)
          (2, 0.0529)
          (3, 0.0529)
          (4, 0.0529)
          (5, 0.0529)
          (6, 0.0529)
          (7, 0.0529)
          (8, 0.0529)
          (9, 0.0529)
        };
        \addplot[thick,color=matisse,mark=square] coordinates {
          (1, 0.0567)
          (2, 0.0558)
          (3, 0.0568)
          (4, 0.0565)
          (5, 0.0598)
          (6, 0.0569)
          (7, 0.0561)
          (8, 0.0576)
          (9, 0.0556)
        };
        \addplot[thick,color=sun_shade,mark=diamond] coordinates {
          (1, 0.0537)
          (2, 0.0539)
          (3, 0.0534)
          (4, 0.0529)
          (5, 0.0572)
          (6, 0.0562)
          (7, 0.0536)
          (8, 0.0547)
          (9, 0.0535)
        };
        \addplot[thick,color=flamingo,mark=*] coordinates {
          (1, 0.0520)
          (2, 0.0538)
          (3, 0.0525)
          (4, 0.0528)
          (5, 0.0531)
          (6, 0.0529)
          (7, 0.0543)
          (8, 0.0565)
          (9, 0.0550)
        };
        
        \nextgroupplot[
        title=Yelp,
        ylabel=\large NDCG@20]
         \addplot[dashed, thick,color=free_speech_aquamarine,mark=triangle*] coordinates {
          (1, 0.0188)
          (2, 0.0188)
          (3, 0.0188)
          (4, 0.0188)
          (5, 0.0188)
          (6, 0.0188)
          (7, 0.0188)
          (8, 0.0188)
          (9, 0.0188)
        };
        \addplot[thick,color=matisse,mark=square] coordinates {
          (1, 0.0207)
          (2, 0.0208)
          (3, 0.0220)
          (4, 0.0219)
          (5, 0.0233)
          (6, 0.0217)
          (7, 0.0222)
          (8, 0.0224)
          (9, 0.0221)
        };
        \addplot[thick,color=sun_shade,mark=diamond] coordinates {
          (1, 0.0210)
          (2, 0.0210)
          (3, 0.0206)
          (4, 0.0211)
          (5, 0.0219)
          (6, 0.0214)
          (7, 0.0210)
          (8, 0.0216)
          (9, 0.0209)
        };
        \addplot[thick,color=flamingo,mark=*] coordinates {
          (1, 0.0205)
          (2, 0.0214)
          (3, 0.0208)
          (4, 0.0211)
          (5, 0.0208)
          (6, 0.0211)
          (7, 0.0216)
          (8, 0.0222)
          (9, 0.0216)
        };

	\end{groupplot}
\node at ($(group c1r1) + (110pt, 100pt)$) {\ref{grouplegend2}};
\end{tikzpicture}
    \caption{Impact of the different augmentations with different \large proportion $\eta$, $\gamma$, and $\beta$ on \large HR@20 and \large NDCG@20. The dash line is the performance of the SASRec method.} 
    \label{fig:Proportion2}
\end{figure}
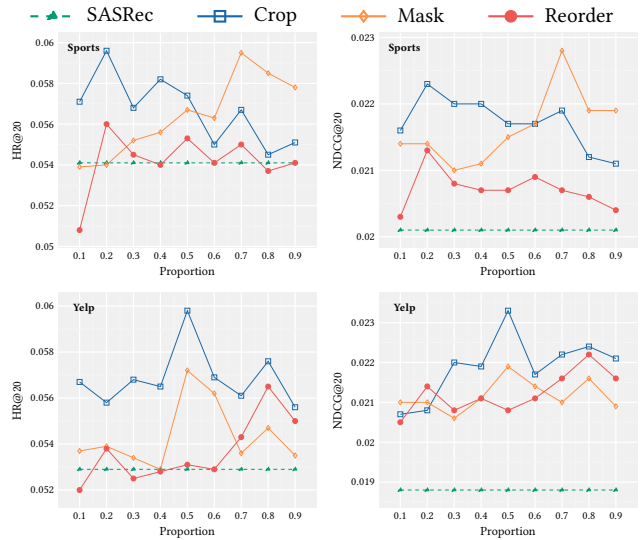

Second, we observe how the proportion rates of different augmentation methods affect the recommendation performance.
Considering the item crop and item mask operators, a general pattern is that the performance peaks at a special proportion rate and then deteriorates if we increase or decrease the rate.
For example, item mask operator peaks at the proportion rate of 0.5 on the Yelp dataset.
It can be explained that when the proportion rate $\gamma$ equals to 0, item mask operator does not function and when $\gamma$ equals to 1.0, the whole user sequence only consists of $[mask]$ items, thus hurting the performance.
Item crop operator acts similarly to item mask operation, except that item crop operator does not function with $\eta=1.0$ and the user sequence is empty with $\eta=0$.

\subsection{Impact of Contrastive Learning Loss (RQ3)}
\pgfplotsset{
axis background/.style={fill=gallery},
grid=both,
  xtick pos=left,
  ytick pos=left,
  tick style={
    major grid style={style=white,line width=1pt},
    minor grid style=bgc,
    draw=none
    },
  minor tick num=1,
  ymajorgrids,
	major grid style={draw=white},
	y axis line style={opacity=0},
	tickwidth=0pt,
}

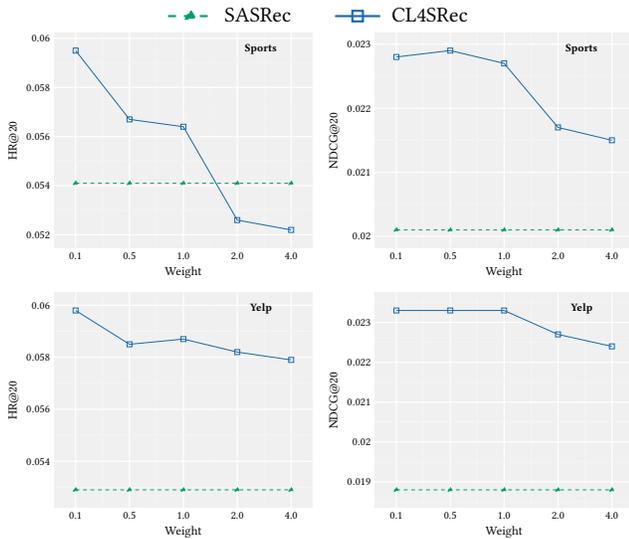
\begin{figure}
\centering
    \begin{tikzpicture}[scale=0.47]
	\begin{groupplot}[
	    group style={group size=2 by 2,
	        horizontal sep = 50pt,
	        vertical sep=4em
	        }, 
	    width=0.5\textwidth,
	    xlabel=\large Weight,
        ylabel=\large HR@20,
        xticklabels={0.1, 0.5, 1.0, 2.0, 4.0},
        xtick={1,2,3,4,5},
        ymajorgrids,
        major grid style={draw=white},
        y axis line style={opacity=0},
        tickwidth=0pt,
        yticklabel style={
        /pgf/number format/fixed,
        /pgf/number format/precision=5
        },
        scaled y ticks=false,
        every axis title/.append style={at={(0.8,0.85)},font=\bfseries}
	    ]
		\nextgroupplot[
		legend style = {
		  font=\small,
          draw=none, 
          fill=none,
          column sep = 2pt, 
          /tikz/every even column/.append style={column sep=5mm},
          legend columns = -1, 
          legend to name = grouplegend},
		title=Sports, 
		]
		
         \addplot[dashed, thick,color=free_speech_aquamarine,mark=triangle*] coordinates {
          (1, 0.0541)
          (2, 0.0541)
          (3, 0.0541)
          (4, 0.0541)
          (5, 0.0541)
        }; \addlegendentry{SASRec}
        \addplot[thick,color=matisse,mark=square] coordinates {
          (1, 0.0595)
          (2, 0.0567)
          (3, 0.0564)
          (4, 0.0526)
          (5, 0.0522)
        };\addlegendentry{CL4SRec}
        
        \nextgroupplot[
        title=Sports,
        ylabel=\large NDCG@20]
         \addplot[dashed, thick,color=free_speech_aquamarine,mark=triangle*] coordinates {
          (1, 0.0201)
          (2, 0.0201)
          (3, 0.0201)
          (4, 0.0201)
          (5, 0.0201)
        };
        \addplot[thick,color=matisse,mark=square] coordinates {
          (1, 0.0228)
          (2, 0.0229)
          (3, 0.0227)
          (4, 0.0217)
          (5, 0.0215)
        };
        
        \nextgroupplot[
        title=Yelp,
        ylabel=\large HR@20]
         \addplot[dashed, thick,color=free_speech_aquamarine,mark=triangle*] coordinates {
          (1, 0.0529)
          (2, 0.0529)
          (3, 0.0529)
          (4, 0.0529)
          (5, 0.0529)
        };
        \addplot[thick,color=matisse,mark=square] coordinates {
          (1, 0.0598)
          (2, 0.0585)
          (3, 0.0587)
          (4, 0.0582)
          (5, 0.0579)
        };
        
        \nextgroupplot[
        title=Yelp,
        ylabel=\large NDCG@20]
         \addplot[dashed, thick,color=free_speech_aquamarine,mark=triangle*] coordinates {
          (1, 0.0188)
          (2, 0.0188)
          (3, 0.0188)
          (4, 0.0188)
          (5, 0.0188)
        };
        \addplot[thick,color=matisse,mark=square] coordinates {
          (1, 0.0233)
          (2, 0.0233)
          (3, 0.0233)
          (4, 0.0227)
          (5, 0.0224)
        };
        
	\end{groupplot}
\node at ($(group c1r1) + (110pt, 100pt)$) {\ref{grouplegend}};
\end{tikzpicture}
    \caption{Performance comparison on CL4SRec w.r.t. different $\lambda$. The dash line is the performance of SASRec. } 
    \label{fig:lambda}
\end{figure}
To answer \textbf{RQ3}, we investigate how the contrastive learning loss of our proposed CL4SRec interacts with the sequential prediction loss.
Specifically, we explore how different $\lambda$ impact the recommendation performance.
We select the best augmentation method with best proportion rate for each dataset according to the results in Section 4.3 and keep other parameters fixed to make a fair comparison.

Figure~\ref{fig:lambda} shows the evaluation results. 
Note that with larger value $\lambda$, $\mathcal{L}_{\mathrm{cl}}$ contributes more heavily in the $\mathcal{L}_{\mathrm{total}}$.
We observe that performance substantially deteriorates when $\lambda$ increases over than certain threshold.
However, with proper $\lambda$, CL4SRec is consistently better than SASRec in all cases.
This observation implies that when contrastive learning loss dominates the learning process, it may decrease the performance on the sequence prediction task.
We will analyze this impact carefully in future work.

\subsection{Ablation Study(RQ4)}
We perform an ablation study on CL4SRec by showing how the augmentation methods and contrastive learning loss affect its performance.
To verify the effectiveness of each component, we conduct experiments on the variant of SASRec, called SASRec$_{\mathrm{aug}}$, which enhances the behavior sequences using our proposed augmentation methods during the training procedure.

\begin{table}
  \caption{Performance of SASRec, SASRec$_{\mathrm{aug}}$, and CL4SRec on the metrics HR@20 and NDCG@20.}
  \centering
  \label{tab:only aug}
  \begin{adjustbox}{max width=\linewidth}
  \begin{tabular}{l lc cc c}
    \toprule
    \multirow{2}{*}{Aug} & \multirow{2}{*}{Method} & \multicolumn{2}{c}{Sports} & \multicolumn{2}{c}{Yelp} \\
     \cmidrule(lr){3-4} \cmidrule(lr){5-6}
     &  & HR@20 & NDCG@20 & HR@20 & NDCG@20\\
    \midrule
    None & SASRec & 0.0541 & 0.0201 & 0.0529 & 0.0188 \\
    \midrule
    \multirow{2}{*}{Crop}
    &SASRec$_{\mathrm{aug}}$ &0.0546 & 0.0205 & 0.0576 & 0.0222 \\
    &CL4SRec & 0.0596 & 0.0223 & 0.0598 & 0.0233 \\
    \midrule
    
    \multirow{2}{*}{Mask}
    &SASRec$_{\mathrm{aug}}$ &0.0513 & 0.0198 & 0.0546 & 0.0206 \\
    &CL4SRec & 0.0595 & 0.0228 & 0.0572 & 0.0219 \\
    \midrule
    
    \multirow{2}{*}{Reorder}
    &SASRec$_{\mathrm{aug}}$ & 0.0547 & 0.0208 & 0.0523 & 0.0203 \\
    &CL4SRec & 0.0560 & 0.0213 & 0.0565 & 0.0222 \\

  \bottomrule
\end{tabular}
\end{adjustbox}
\end{table}

Table \ref{tab:only aug} shows the results of SASRec, SASRec$_{\mathrm{aug}}$, and CL4SRec.
We observe a few trends in this table.
On the one hand, we find that SASRec$_{\mathrm{aug}}$ outperforms SASRec in all datasets with almost all augmentations. 
This indicates that our augmentation methods are very useful for basic models by adding random noises.
On the other hand, with our proposed contrastive learning loss, CL4SRec consistently outperforms SASRec$_{\mathrm{aug}}$ in all datasets with all augmentations.
It verifies the effectiveness of our contrastive learning component for providing significant self-supervised signals, as illustrated above.

\subsection{Quality of User Representations (RQ5)}
As mentioned above, CL4SRec can infer better representations in the user behavior sequence level compared with other state-of-the-art baselines.
In this subsection, we attempt to verify whether CL4SRec can really cope with this problem.
We utilize the friend relations provided by Yelp dataset to explore whether users are closer to each other in the latent space if they are friends.
Cosine function is applied to measure the similarity.
Note that we do not use this kind of information during the training procedure.

\begin{figure}
\pgfplotsset{
axis background/.style={fill=gallery},
grid=both,
  xtick pos=left,
  ytick pos=left,
  tick style={
    major grid style={style=white, line width=1pt},
    minor grid style=white,
    draw=none,
  },
  minor tick num=1,
}
\centering
\resizebox{\linewidth}{!}{
    \begin{tikzpicture}
      \begin{axis}[
        xlabel=\huge Example,
        ylabel=\huge Count,
        ymin=0, ymax=220, ytick distance=20,
        xmax=1, xmin=-1, xtick distance=0.1,
        x tick label style={
        font=\small,
        scaled ticks=false,
        /pgf/number format/fixed,
        /pgf/number format/precision=2},
        y tick label style={
        font=\normalsize,
        },
        y label style={
         yshift=3mm
        },
        x label style={
         yshift=-2mm
        },
        legend style={
          legend image post style={scale=2},
          font=\Large,
          fill=gallery,
          draw=none,
          draw opacity=0,
          legend cell align=left,
          legend columns=1,
          at={(0.1, 0.97)},
          anchor=north,
          row sep=2mm,
          column sep=2mm,
          /tikz/every even column/.append style={column sep=1mm}
        },
        width=\textwidth,
        height=0.618\textwidth,
        ybar=0pt, %
        bar shift=0pt,
        bar width=10pt,
        ymajorgrids,
        major grid style={draw=white},
        y axis line style={opacity=0},
        tickwidth=0pt,
        legend entries = {cl, sas}
        ]
        \addplot[draw=none, fill=matisse, opacity=0.5] coordinates {
        (-0.50, 2)
        (-0.45, 2)
        (-0.40, 2)
        (-0.35, 2)
        (-0.30, 5)
        (-0.25, 2)
        (-0.20, 10)
        (-0.15, 20)
        (-0.10, 11)
        (-0.05, 13)
        (0.00, 20)
        (0.05, 15)
        (0.10, 27)
        (0.15, 28)
        (0.20, 30)
        (0.25, 40)
        (0.30, 70)
        (0.35, 74)
        (0.40, 79)
        (0.45, 103)
        (0.50, 111)
        (0.55, 151)
        (0.60, 157)
        (0.65, 191)
        (0.70, 182)
        (0.75, 200)
        (0.80, 207)
        (0.85, 166)
        (0.90, 114)
        (0.95, 26)
          };        
        \addplot[draw=none, fill=flamingo, opacity=0.5] coordinates {
        (-0.35, 1)
        (-0.30, 6)
        (-0.25, 3)
        (-0.20, 12)
        (-0.15, 9)
        (-0.10, 19)
        (-0.05, 33)
        (0.00, 26)
        (0.05, 34)
        (0.10, 37)
        (0.15, 45)
        (0.20, 37)
        (0.25, 57)
        (0.30, 75)
        (0.35, 107)
        (0.40, 147)
        (0.45, 157)
        (0.50, 186)
        (0.55, 212)
        (0.60, 211)
        (0.65, 170)
        (0.70, 177)
        (0.75, 144)
        (0.80, 98)
        (0.85, 39)
        (0.90, 7)
        (0.95, 11)
          };
      \end{axis}
      
    \end{tikzpicture}}
    \caption{Comparison of cosine similarity distribution between friends in the Yelp dataset. Each bar at the range $[x, x+0.05)$ demonstrates the count of users whose similarity score with her/his friends falls in this range.}
  \label{user}
\end{figure}
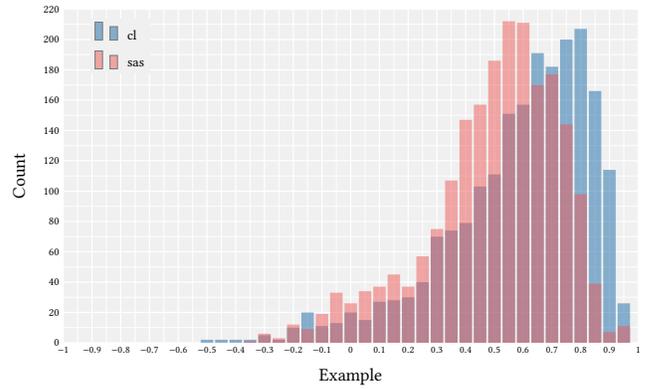

Figure~\ref{user} shows the evaluation results.
The average cosine similarity is 0.5198 and 0.6100 for SASRec and CL4SRec, respectively.
We can observe that the representations of similar users inferred by CL4SRec are much closer in the latent space than SASRec.
This demonstrates that CL4Rec can really capture user preference hidden in her/his interaction sequences, thus deriving better representations in the user behavior sequence level.

\section{Conclusion and Future Work}
In this paper, we propose a novel model called Contrastive Learning for Sequential Recommendation (CL4SRec), which can learn the effective user representation model only with interaction information.
It is equipped with the contrastive learning framework to extract self-supervised signals just from raw data and utilize them to improve the base model.
In addition, we propose three data augmentation approaches to construct contrastive tasks and exploit the effects of the composition of different augmentations. 
The proposed method is verified on four public datasets. 
Extensive experiment results show that our CL4SRec achieves significant improvements and outperforms the start-of-the-art baselines.

\bibliographystyle{ACM-Reference-Format}
\bibliography{CL4Rec}

\end{document}